\documentclass[twocolumn,pra,superscriptaddress]{revtex4-1}
\usepackage{mathtools}
\usepackage{sfmath}
\usepackage{braket}

\usepackage[colorlinks,
linkcolor=blue,
anchorcolor=blue,
citecolor=blue,
filecolor=blue,
menucolor=blue,
runcolor=blue,
urlcolor=blue,
frenchlinks=blue]{hyperref}

\makeatletter

\newcommand{\Rmnum}[1]{\expandafter\@slowromancap\romannumeral #1@}
\usepackage[utf8]{inputenc}

\begin{document}

\title{Quantum phase transition in a non-Hermitian XY spin chain with global complex transverse field}

\author{Yu-Guo Liu}
\affiliation{School of Physics, Nanjing University, Nanjing 210093, China}
\affiliation{National Laboratory of Solid State Microstructures, Collaborative Innovation Center of Advanced Microstructures,  Nanjing University, Nanjing 210093, China}

\author{Lu Xu}
\affiliation{School of Physics, Nanjing University, Nanjing 210093, China}
\affiliation{National Laboratory of Solid State Microstructures, Collaborative Innovation Center of Advanced Microstructures,  Nanjing University, Nanjing 210093, China}

\author{Zhi Li}
\email{lizhiphys@126.com}
\affiliation{Guangdong Provincial Key Laboratory of Quantum Engineering and Quantum Materials, SPTE, South China Normal University, Guangzhou 510006, China}
\affiliation{GPETR Center for Quantum Precision Measurement, South China Normal University, Guangzhou 510006, China}
\affiliation{Guangdong-Hong Kong Joint Laboratory of Quantum Matter, Frontier Research Institute for Physics, South China Normal University, Guangzhou 510006, China}

\date{\today}

\begin{abstract}
In this work, we investigate the quantum phase transition in a non-Hermitian XY spin chain. The phase diagram shows that the critical points of Ising phase transition expand into a critical transition zone after introducing a non-Hermitian effect. By analyzing the non-Hermitian gap and long-range correlation function, one can distinguish different phases by means of different gap features and decay properties of correlation function, a tricky problem in traditional XY model. Furthermore, the results reveal the relationship among different regions of the phase diagram, non-Hermitian energy gap and long-range correlation function.
\end{abstract}


\maketitle


\section{Introduction}

Quantum XY spin chain is a textbook model in exploring quantum magnetism and quantum phase transitions (QPTs)~\cite{Sachdev2011}. It is extended from one-dimensional transverse field Ising model by adding the spin-spin  interaction along another direction. There are two types of QPTs in the XY model, i.e., Ising phase transition~\cite{Pfeuty1970,McCoy1971} and anisotropic phase transition~\cite{McCoy1971,Daniel1961,Vidal2004}, which can be characterized by different critical behaviors. Since most XY models and their derivatives can be analytically solved by Jordan-Wigner transformation~\cite{Jordan1928} or numerically solved by the renormalization group method~\cite{Wilson1983,Doniach1978,Pfeuty1979,Langari2008}, XY model has attracted wide attention and fruitful research results have been obtained during the past decades~\cite{Ming2001,Vidal2004,Vahedi2018,Song2017,Song2015,Ross1999,Tong2010,Tong2013,Ugo2017,Gao2017,Zhu2006,Lipeng2020}. On the other hand, great attention has been paid to the non-Hermitian systems as experimental techniques develop rapidly in recent years. Not only can non-Hermitian systems be readily realized through multiple existing table-top experimental platforms (such as cold atomic system~\cite{luole2019,Jiangbin2020}, optical system~\cite{YangLan2017,YangLan2014,Rechtsman2019}, nitrogen-vacancy center~\cite{Jiangfeng2019}, etc.), but they can also trigger many novel physical phenomena (such as real eigenvalues with parity-time ($\mathcal{PT}$) symmetry~\cite{Bender1998}, non-Hermitian skin effect~\cite{YaoShunyu2018,WangZhong2018}, new topological properties corresponding to exceptional points (EPs)~\cite{Danwei2020,Hepeng2020,Fuliang2018,Flore2019,Ghatak2019,Masatoshi2019,Heyan2020} and disorders~\cite{Zhang2020,Chuanwei2020,Tang2020,Jiang2019}). What would happen if QPTs meet non-Hermitian effects?

In general, QPTs fall into two broad categories: traditional QPTs~\cite{Sachdev2011,Lee2014,Yamamoto2019,Kou2020} and topological QPTs~\cite{ShouCheng2011,Kane2010,Zhang2019,Tan2019,Wang2019,Zhu2013}. The former can be depicted by local order parameters, while the latter are characterized by the global topological invariants. The non-Hermiticity gives rise to a brand-new phase transition, so called as non-Hermitian QPTs, characterized by energy spectrum. The new phase transition is closely related to particular symmetries, for instance, $\mathcal{PT}$ symmetry and intrinsic rotation-time-reversal ($\mathcal{RT}$) symmetry~\cite{ZhangXZ2013,WangCan2020,Lee2014,LiC2014,ZhangKL2020}. The system features pure real energy spectrum in symmetry-preserving region, whereas it possesses complex energy spectrum in the region of broken symmetry~\cite{ZhangXZ2013,WangCan2020}. Recent years have witnessed extensive investigation of the non-Hermitian QPTs and there are also some works for the spin system in the complex field or with the non-Hermitian interactions~\cite{Yoshihiro2020,Nishiyama2020,Zhang2013,Lipeng2020}. However, the research remains inadequate on the influence of non-Hermiticity on the traditional QPTs in spin system. Recently, there are some works that investigated non-Hermitian quantum criticality in real-spectrum region by biorthogonal fidelity susceptibility~\cite{Kou2020,Tzeng2021}. But it is still an open question that how the non-Hermiticiy affects the traditional QPT and quantum magnetism in the complex-spectrum region or the system without $\mathcal{PT}$ or $\mathcal{RT}$ symmetry.

This paper is devoted to the research on the traditional QPTs in the presence of non-Hermitian effects, which are derived from global complex transverse field. First, we define the non-Hermitian ground state and energy gap. Based on characteristics of the energy gap, phase diagram of the system can be divided into three regions, which are corresponding to pure real gap, pure imaginary gap and complex gap, respectively. Besides, all gapless points form an exceptional ring. By studying the non-analyticity of the ground state energy density, we find that the phase transition which occurs on the exceptional ring is actually a second-order phase transition. Moreover, through the exact solution and numerical fitting, we investigate the long-range correlation function (LRCF) in different regions of the phase diagram. We discover the corresponding relations among the QPT, non-Hermitian energy gap and LRCF, i.e., ferromagnetic phase (critical transition zone, paramagnetic phase) corresponds to pure real (pure imaginary, complex) gap, whose LRCF features no decay (polynomial decay, exponential decay).

The paper is organized as follows. In Sec.\,2, the model, the exact solution and its non-Hermitain ground state are provided. In Sec.\,3, we define the non-Hermitian gap, draw the phase diagram and point out that the QPT on the exceptional ring is the second-order phase transition. In Sec.\,4, we study the LRCF and analyze its decay behavior. Sec.\,5 is the conclusion.

\section{Model and non-Hermitian ground state}
The Hamiltonian of a non-Hermitian quantum XY spin chain with a global complex transverse field reads
\begin{equation}\label{H}
\begin{split}
H=&-\sum_{j=-N/2}^{N/2-1} \left( \frac{J+\gamma}{2}\sigma_j^x \sigma_{j+1}^x+\frac{J-\gamma}{2}\sigma_j^y \sigma_{j+1}^y \right)\\
  &-\sum_{j=-N/2}^{N/2} \left(\lambda \sigma_j^z+\frac{i\Gamma}{2}\sigma_j^{u} \right)\;,
\end{split}
\end{equation}
where $N$ denotes the total sites number of the chain. Since $N$ is large enough, $N/2$ could be considered a ``decent half" no matter $N$ is even or odd. The Hilbert space on each site has a set of basis vectors of two spin states, i.e., $\ket{\uparrow}$ and $\ket{\downarrow}$. $\sigma_j^x$, $\sigma_j^y$ and $\sigma_j^z$ are the Pauli matrices for spin $j$ and $\sigma^{u}$ denotes the matrix $\bigl[\begin{smallmatrix} 1 & 0\\ 0 & 0 \end{smallmatrix} \bigr]$, which corresponds to the gain (\begin{small}$\Gamma<0$\end{small}) or loss (\begin{small}$\Gamma>0$\end{small}) of $\ket{\uparrow}$ with the rate of $\Gamma/2$. The system can be reduced to Hermitian XY model when $\Gamma=0$. $\lambda$ parameterizes transverse magnetic field along $z$ direction. $J$ ($\gamma$) represents the isotropic (anisotropic) nearest neighboring interaction strength between $x$ and $y$ directions. All the parameters are real numbers. The sign of $J$ determines that the system is characterized by ferromagnetic or antiferromagnetic chains. Since we hereby just concentrate on the ferromagnetism case, $J>0$ and $\gamma/J \in [-1,1]$ are taken into consideration. Besides, $J$ as the system character parameter is set as unit 1 later in this paper.

We can diagonalize the Hamiltonian in Eq.\,(\ref{H}) by three steps. First, we map spin operators onto spinless Fermi operators by Jordan-Wigner transformation, i.e.,
\begin{equation}\label{FH}
\begin{split}
&\sigma_j=e^{-i\pi\sum_{l<j}c_l^\dag c_l} c_j\;,\ \ \ \ \ \sigma_j^\dag=c_j^\dag e^{i\pi\sum_{l<j}c_l^\dag c_l}\;,\\
&c_j=e^{i\pi\sum_{l<j}\sigma_l^\dag \sigma_l} \sigma_j\;,\ \ \ \ \ \ c_j^\dag=\sigma_j^\dag e^{-i\pi\sum_{l<j}\sigma_l^\dag \sigma_l}\;,
\end{split}
\end{equation}
where $\sigma=\bigl[\begin{smallmatrix} 0 & 0\\ 1 & 0 \end{smallmatrix}\bigr]$, $\sigma^\dag=\bigl[\begin{smallmatrix} 0 & 1\\ 0 & 0 \end{smallmatrix}\bigr]$, and $c_j$ ($c_j^\dag$) denotes the annihilation (creation) operators of the fermion on the $jth$ site. Then, we can obtain the Hamiltonian in the spinless fermion representation as
\begin{equation}\label{FR}
\begin{split}
H=&-\sum_{j=-N/2}^{N/2-1} \left(J c_j^\dag c_{j+1} + \gamma c_j^\dag c_{j+1}^\dag + h.c. \right)\\
  &+\sum_{j=-N/2}^{N/2} \left[\lambda \mathbf{1}_j - (2\lambda + i\frac{\Gamma}{2}) c_j^\dag c_j \right]\;.
\end{split}
\end{equation}
Second, we do Fourier transformation with $c_k=\frac{e^{i\pi/4}}{\sqrt{N}}\sum_j\text{exp}(-ikj)c_j$.
The corresponding Hamiltonian in momentum space reads,
\begin{equation}\label{PR}
H=\sum_{k=0}^{\pi} \left[-d_0(k) + (c_k^\dag\ \  c_{-k}) \mathcal{H}(k) \left(\begin{matrix} c_k\\ c_{-k}^\dag\\ \end{matrix} \right) \right]\;.
\end{equation}
$\mathcal{H}(k)$ is the BdG Hamiltonian,
\begin{equation}\label{BH}
\mathcal{H}(k)= d_x(k)\tau_x + d_y(k)\tau_y + d_z(k)\tau_z\;,
\end{equation}
where the $\tau_i$ ($i=x,y,z$) is the Pauli matrix of pseudo spin and
\begin{equation}\label{d}
\begin{split}
&d_0(k)= 2J\cos{k} +i\Gamma/2\;,\\
&d_x(k)=2\gamma\sin{k}\;,\\
&d_y(k)=0\;,\\
&d_z(k)=-(2\lambda + 2J\cos{k} + i\Gamma/2)\;.
\end{split}
\end{equation}
Third, we use non-Hermitian Bogoliubov transformation to complete diagonalization of Hamiltonian in Eq.\,(\ref{PR}) following the method outlined in Ref.~\cite{Lee2014}. In Ref.~\cite{Lee2014}, the authors investigate nonequilibrium steady state, which has minimum imaginary part of eigenvalue. Here, we focus on the state with minimum real part of energy, which is defined as the non-Hermitian ground state in this paper. Thus, our definitions of the $u$, $v$ factors of Bogoliubov transformation are different from that in Ref.~\cite{Lee2014}.

We define the annihilation ($\alpha$) and creation ($\bar{\alpha}$) operators of non-Hermitian Bogoliubov quasiparticles as
\begin{equation}
\begin{split}
&\alpha_k=u_k c_k + v_k c_{-k}^\dag\;, \ \ \ \ \alpha_{-k}=-v_k c_k^\dag + u_k c_{-k}\;,\\
&\bar{\alpha}_k=u_k c_k^\dag + v_k c_{-k}\;, \ \ \ \ \bar{\alpha}_{-k}=-v_k c_k + u_k c_{-k}^\dag\;.
\end{split}
\end{equation}
The $u$, $v$ factors are defined as $u_k=\cos{(\theta_k/2)}$ and $v_k=\sin{(\theta_k/2)}$ with the relations $\sin{\theta_k}=d_x(k)/E_k$, $\cos{\theta_k}=d_z(k)/E_k$, where
\begin{equation}\label{Ek}
E_k=\sqrt{d_x(k)^2+d_z(k)^2}\;.
\end{equation}
Then, the Hamiltonian is diagonalized as
\begin{equation}\label{BR}
H=-\sum_{k=0}^{\pi} \bigl( d_0(k)+E_k \bigr) + \sum_{k=0}^{\pi} E_k \bigl( \bar{\alpha}_k \alpha_k + \bar{\alpha}_{-k} \alpha_{-k} \bigr)\;.
\end{equation}
Note that, since $u$, $v$ are complex numbers, $\alpha_k^\dag \ne \bar{\alpha}_k$. The fermionic commutation relation is still held, i.e., $\{\bar{\alpha}_k, \alpha_{k'}\}=\delta_{kk'}$ and $\{\alpha_k,\alpha_{k'}\}=\{\bar{\alpha}_k,\bar{\alpha}_{k'}\}=0$. As for the excited state energy $E_k$, we choose the branch cut of square root along the negative x-axis so that the real part of $E_k$ is always nonnegative. Therefore, we can define the non-Hermitian ground state $|G\rangle$ as the state with the minimum real part of energy. $|G\rangle$ can be obtained by the equation $\alpha_k|G\rangle=\alpha_{-k}|G\rangle=0$, i.e.,
\begin{equation}\label{G}
|G\rangle=\prod_{k=0}^{\pi} \frac{1}{\sqrt{\mathcal{N}}} \bigl( u_k - v_k c_k^\dag c_{-k}^\dag \bigr)|0\rangle\;,
\end{equation}
where $|0\rangle$ is the vacuum state, $\mathcal{N}=\prod_{k=0}^{\pi}(|u_k|^2+|v_k|^2)$ and the ground state energy $E_G$ is given as
\begin{equation}\label{EG}
E_G=-\sum_{k=0}^{\pi} \bigl(  d_0(k) + E_k \bigr)\;.
\end{equation}
\section{non-Hermitian gap and phase diagram}
Due to the nature of non-Hermiticity, the excited state $\bar{\alpha}_{\pm k}|0\rangle$ has a complex energy $E_k$ described in Eq.\,(\ref{Ek}). Notably, the $E_k$ have nonnegative real part by the negative-x-axis branch cut of square root, so we can find a momentum $k_m$ to corresponds to the least real part of excitation energy, i.e., minimum value of ${\rm Re}[E(k_m)]$. We define the minimum value as the non-Hermitian gap, which is denote as $\Delta$. In Hermitian systems, gapless points are usually the boundary of different phases. A natural question is whether the $\Delta$ will also become a landmark in non-Hermitian systems. Therefore, we divide the parameter space $(\lambda, \gamma, \Gamma)$ of the Hamiltonian (\ref{H}) by $\Delta$.

\begin{figure}[htbp]\centering
\includegraphics[width=9cm]{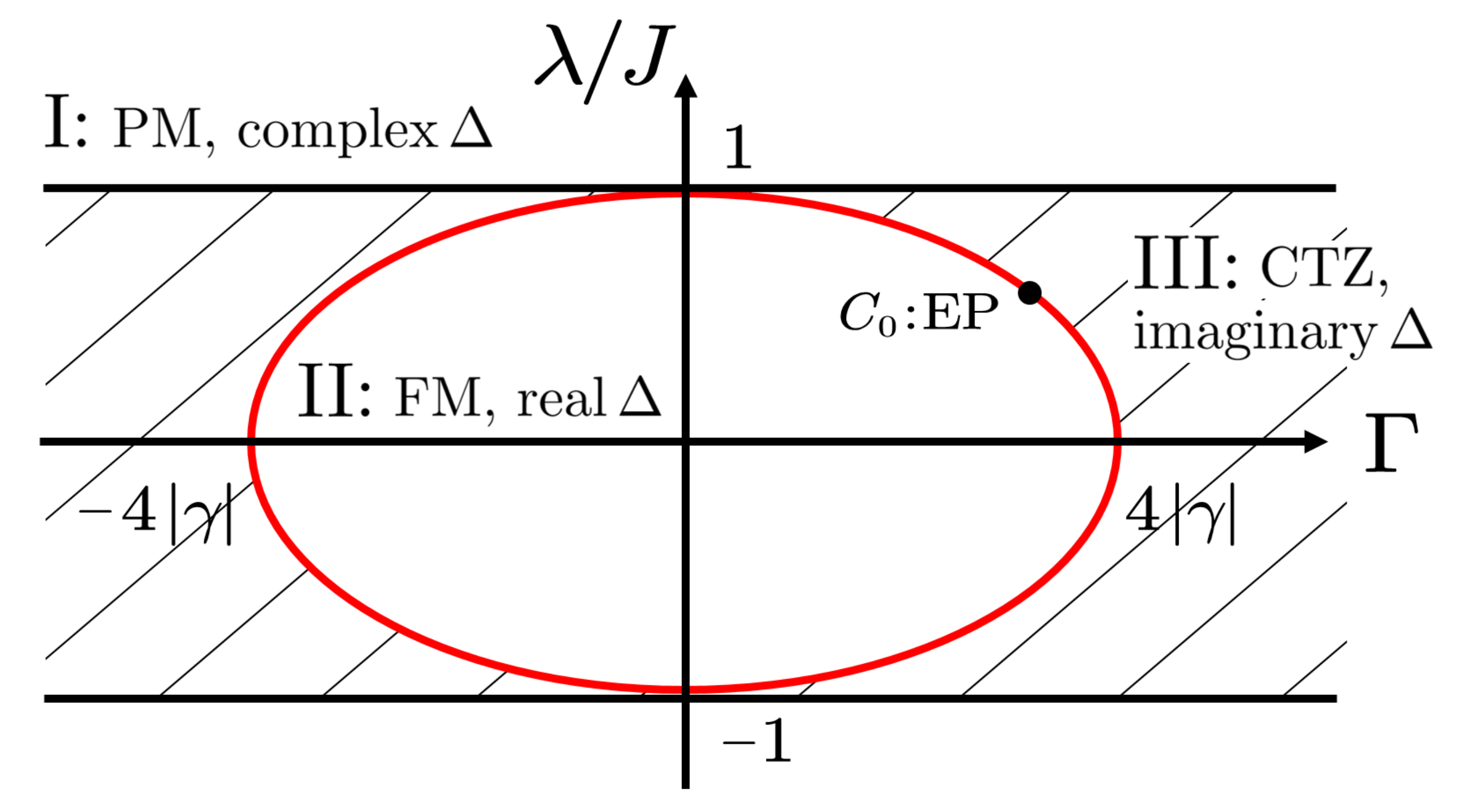}
 \caption{Phase diagram of the non-Hermitian XY model. The red line denotes the exceptional ring, which corresponds elliptic equation (Eq.~(\ref{ring})). In region \Rmnum{1} ($|\lambda/J|>1$), it is in the paramagnetic (PM) phase with a complex non-Hermitian gap. In region \Rmnum{2}, it is in the ferromagnetic (FM) phase with a pure real gap. In region \Rmnum{3}, the shaded area, the non-Hermitian gap is a pure imaginary number. We call region \Rmnum{3} as critical transition zone (CTZ).}\label{Pdiagram}
\end{figure}

First, by solving the equation $E_k=0$, one can get
\begin{subequations}\label{EP}
\begin{eqnarray}
 	&&\frac{\lambda^2}{J^2}+\frac{\Gamma^2}{(4\gamma)^2}=1\;,\label{ring}\\
 	&&\cos k_m=-\frac{\lambda}{J}\;.\label{cosk}
 \end{eqnarray}
 \end{subequations}
The Eq.\,(\ref{ring}) depicts an elliptical exceptional ring, and the Eq.\,(\ref{cosk}) implies a limitation $|\lambda|\le J$. Therefore, the system can be separated into three regions as shown in Fig.~\ref{Pdiagram}.

Region \Rmnum{1}: $|\lambda|>J$. This region is characterized by a complex gap with ${\rm Re}[\Delta]>0$ and $|{\rm Im}[\Delta]|=|\Gamma|/2$, where the ${\rm Re}[\Delta]$ and ${\rm Im}[\Delta]$ represent the real and imaginary part of complex gap $\Delta$, respectively. We take $\lambda>J$ as an example. Under this condition, $k_m=\pi$ and $\Delta=2\lambda-2J+i\Gamma/2$.

Region \Rmnum{2}: inside the exceptional ring. This region has pure real gap, i.e., ${\rm Re}[\Delta]>0$ and ${\rm Im}[\Delta]=0$. In this region, $k_m=\arccos(-\frac{\lambda}{J})$ and $\Delta=\sqrt{4\gamma^2(1-\lambda^2/J^2)-\Gamma^2/4}$.

Region \Rmnum{3}: outside the exceptional ring and $|\lambda|<J$. The gap is pure imaginary number (${\rm Re}[\Delta]=0$ and ${\rm Im}[\Delta]\ne0$). In this region, we still have $k_m=\arccos(-\frac{\lambda}{J})$. However, $\Delta$ becomes a pure imaginary number because the non-Hermitian strength $\Gamma$ is large enough to be dominant, then $\Delta=\pm i\sqrt{\Gamma^2/4-4\gamma^2(1-\lambda^2/J^2)}$.

\begin{figure}[htbp]\centering
\includegraphics[width=8.6cm]{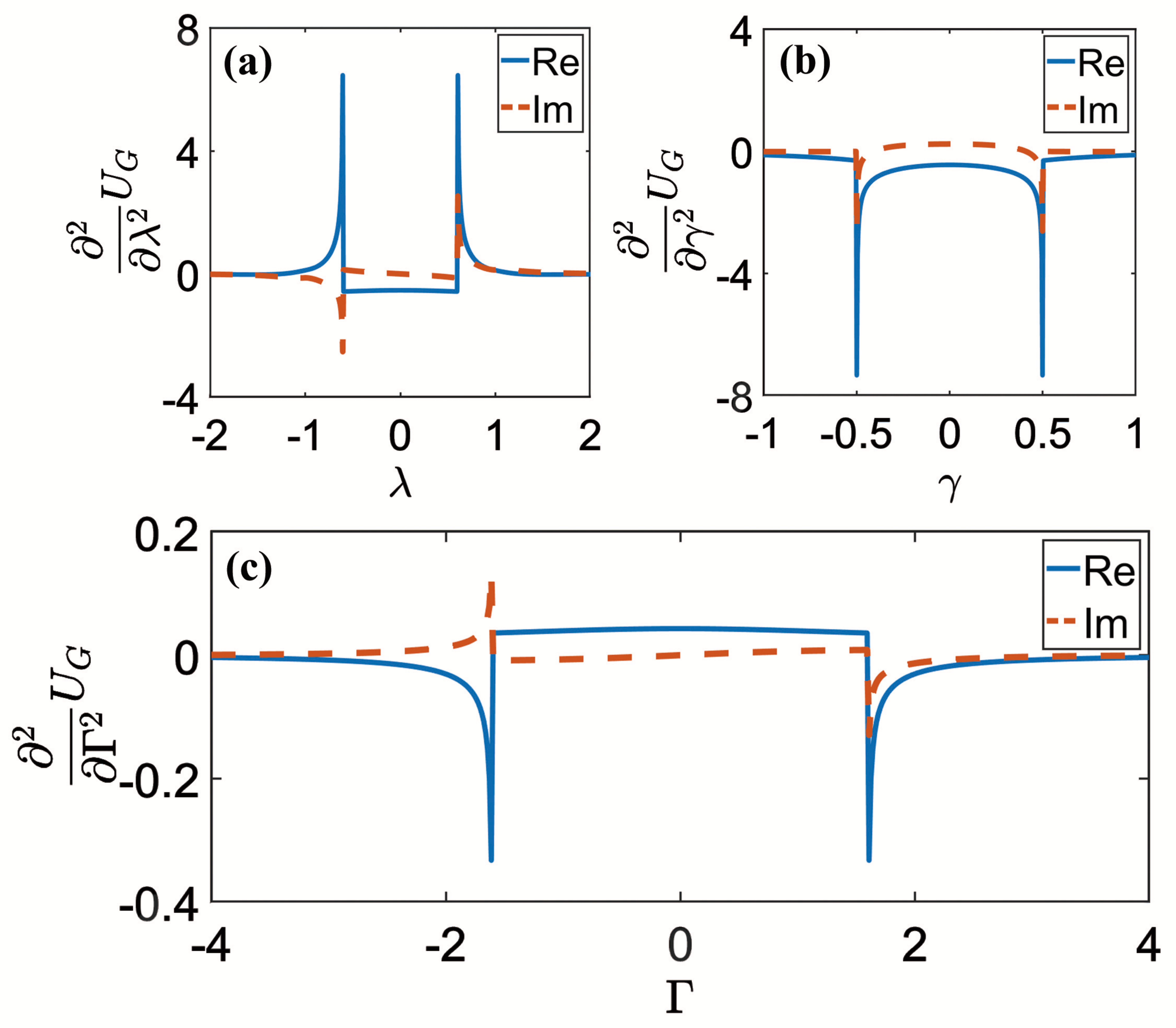}
 \caption{Second derivatives of ground state energy density versus (a) $\lambda$ with $(\gamma,\,\Gamma)=(0.5,\,1.6)$, (b) $\gamma$ with $(\lambda,\,\Gamma)=(0.6,\,1.6)$ and (c) $\Gamma$ with $(\lambda,\,\gamma)=(0.6,\,0.5)$. The blue solid (red dashed) line represent the real (imaginary) part of the derivatives. Throughout, we set J = 1.}\label{D2}
\end{figure}

Second, we investigate non-analyticity of ground state energy density $U_G$, which implies a QPT at zero temperature. Under the thermodynamic limit, we have
\begin{equation}\label{UG}
U_G=\lim_{N \to \infty} E_G/N = -\frac{1}{2\pi} \int_0^{\pi} (d_0(k) + E_k) \mathrm{d}k\;.
\end{equation}
Moreover, the corresponding first-order derivatives reads
\begin{subequations}
\begin{eqnarray}
&&\frac{\partial U_G}{\partial \lambda}=\frac{-1}{\pi} \int_0^{\pi} \cos \theta_k \mathrm{d}k\;, \\
&&\frac{\partial U_G}{\partial \gamma}=\frac{-1}{\pi} \int_0^{\pi} \sin k \sin \theta_k \mathrm{d}k\;,\\
&&\frac{\partial U_G}{\partial \Gamma}=\frac{-i}{4}-\frac{1}{4\pi} \int_0^{\pi} \cos \theta_k \mathrm{d}k\;,
\end{eqnarray}
\end{subequations}
and the second-order derivatives can be obtained as
\begin{subequations}
\begin{eqnarray}
&&\frac{\partial^2 U_G}{\partial \lambda^2}=\frac{-2}{\pi} \int_0^{\pi} \frac{\sin^2 \theta_k}{E_k} \mathrm{d}k\;,\\
&&\frac{\partial^2 U_G}{\partial \gamma^2}=\frac{-2}{\pi} \int_0^{\pi} \frac{\sin^2 k \cos^2 \theta_k}{E_k} \mathrm{d}k\;,\\
&&\frac{\partial^2 U_G}{\partial \Gamma^2}=\frac{1}{8\pi} \int_0^{\pi} \frac{\sin^2 \theta_k}{E_k} \mathrm{d}k\;.
\end{eqnarray}
\end{subequations}
By numerical calculations, we find that $U_G$ and its first derivatives are continuous while there are divergent points in the second derivatives. The results of second derivatives are plotted in Fig.~\ref{D2}. It is obvious that the second derivatives diverge at the exceptional point $C_0$, which implies that a second-order QPT occurs on the exceptional ring.


\section{Long-Range correlation function in different phase}
The LRCFs are the order parameters of magnetic system in Hermitian spin models. For example, in ferromagnetic phase, LRCF $C_{xx}(r)=\langle G| \sigma_0^x \sigma_r^x |G \rangle$ or $C_{yy}(r)=\langle G| \sigma_0^y \sigma_r^y |G \rangle$ is a non-zero constant, where the parameter $r$ represents the distance between two spins. In paramagnetic phase, LRCF decays exponentially with the increase of $r$. At critical points, correlation functions have asymptotic behavior of polynomial decay, i.e. $C_{xx}(r) \propto r^{2-d-\eta} $, where $d$ and $\eta$ denote the spatial dimension and the critical exponent, respectively. $\eta$ varies in different universality classes. As for XY spin chain, $\eta$ is $5/4$ for the Ising phase transition at $\lambda/J=1$ and $3/2$ for the anisotropic phase transition at $\gamma=0$~\cite{Tong2010,Ross1999}.

In our model, $C_{xx}(r)$ can be calculated through the Pfaffian of a $2r\times2r$ skew symmetric matrix $\rm M$ ($\rm M^\mathrm{T}=-M$)~\cite{Lee2014}
\begin{equation} \label{Cxxr}
C_{xx}(r)= {\rm pf}\bigl( {\rm M}_{2r\times2r}\bigr)= {\rm pf}\left(\begin{matrix}M_{11} &M_{12} \\ -M^\mathrm{T}_{12} &M_{22}\\  \end{matrix} \right)\;,\\
\end{equation}
where the ``~pf~'' means the Pfaffian of a matrix. The numerical calculation codes for Pfaffian are from ``Algorithm 923''~\cite{Wimmer2012}. The elements of matrix $\rm M$ are as following:
\begin{scriptsize}
\begin{equation} \label{M11}
\begin{split}
&\mbox{\normalsize{$M_{11}=$}}\\
&\left(
         \begin{matrix}
         0 &\langle B_0B_1 \rangle &\langle B_0B_2 \rangle &\cdots &\langle B_0B_{r-1} \rangle \\
         -\langle B_0B_1 \rangle &0 &\langle B_1B_2 \rangle &\cdots &\langle B_1B_{r-1}\rangle\\
         -\langle B_0B_2 \rangle&-\langle B_1B_2 \rangle &0 &\cdots &\langle B_2B_{r-1}\rangle\\
         \vdots &\vdots &\vdots &\ddots &\vdots\\
         -\langle B_0B_{r-2} \rangle&-\langle B_1B_{r-2}\rangle &\cdots &\ \ \ \ \ 0 &\langle B_{r-2}B_{r-1} \rangle \\
         -\langle B_0B_{r-1} \rangle&-\langle B_1B_{r-1}\rangle &\cdots &-\langle B_{r-2}B_{r-1} \rangle &0\\
         \end{matrix}
\right),\\
\end{split}
\end{equation}
\end{scriptsize}

\begin{scriptsize}
\begin{equation} \label{M22}
\begin{split}
&\mbox{\normalsize{$M_{22}=$}}\\
&\left(
         \begin{matrix}
         0 &\langle A_1A_2 \rangle &\langle A_1A_3 \rangle &\cdots &\langle A_1A_r \rangle \\
         -\langle A_1A_2 \rangle &0 &\langle A_2A_3 \rangle &\cdots &\langle A_2A_r\rangle\\
         -\langle A_1A_3 \rangle&-\langle A_2A_3\rangle&0 &\cdots &\langle A_3A_r\rangle\\
         \vdots &\vdots &\vdots &\ddots &\vdots\\
         -\langle A_1A_{r-1} \rangle&-\langle A_2A_{r-1}\rangle &\cdots &\ \ \ \ \ 0 & \langle A_{r-1}A_r \rangle \\
         -\langle A_1A_{r} \rangle&-\langle A_2A_{r}\rangle &\cdots  &-\langle A_{r-1}A_r \rangle &0\\
         \end{matrix}
\right),\\
\end{split}
\end{equation}
\end{scriptsize}

\begin{scriptsize}
\begin{equation} \label{M12}
\begin{split}
& \mbox{\normalsize{$M_{12}=$}}\left(
         \begin{matrix}
         \langle B_0A_1 \rangle &\langle B_0A_2 \rangle &\cdots &\langle B_0A_{r} \rangle\\ \\
         \langle B_1A_1 \rangle &\langle B_1A_2 \rangle &\cdots &\langle B_1A_{r}\rangle\\
         \vdots &\vdots &\ &\vdots\\
         \langle B_{r-2}A_1 \rangle &\langle B_{r-2}A_2 \rangle &\cdots &\langle B_{r-2}A_{r}\rangle\\
         \langle B_{r-1}A_1 \rangle &\langle B_{r-1}A_2 \rangle &\cdots &\langle B_{r-1}A_{r}\rangle\\
         \end{matrix}
\right).
\end{split}
\end{equation}
\end{scriptsize}
The pair contractions for $A_m$ and $B_n$ are
\begin{small}
\begin{equation} \label{AB}
\begin{split}
\langle A_mA_n \rangle &=\delta_{mn}+\frac{1}{\pi}\int_0^\pi \mathrm{d}k \sin k(n-m) \left( \frac{u_k v^*_k - u^*_k v_k}{|u_k|^2+|v_k|^2} \right)\;,\\
\langle B_mB_n \rangle &=-\delta_{mn}+\frac{1}{\pi}\int_0^\pi \mathrm{d}k \sin k(n-m) \left( \frac{u_k v^*_k - u^*_k v_k}{|u_k|^2+|v_k|^2} \right)\;,\\
\langle B_mA_n \rangle &=-\langle A_nB_m \rangle\;\\
                       &=-\frac{1}{\pi}\int_0^\pi \mathrm{d}k \cos k(n-m) \left( \frac{|u_k|^2-|v_k|^2}{|u_k|^2+|v_k|^2} \right)\\
                       &\ +\frac{1}{\pi}\int_0^\pi \mathrm{d}k \sin k(n-m) \left( \frac{u_k v^*_k + u^*_k v_k}{|u_k|^2+|v_k|^2} \right)\;.
\end{split}
\end{equation}
\end{small}
It needs to be mentioned that in some works of non-Hermitian systems, the average value of an observable quantity is defined on the biorthogonal bases~\cite{Brody2013,Walker1972}. Here, we do not use the biorthogonal bases because we think it is not easy to measure the biorthogonal average value. The left and right eigenstate are not the same quantum state, thus before measuring the biorthogonal average value, one has to prepare two states, then finds a way to link the two states and the observable quantity together. These processes are inconvenient for experiments. Thus, in this work, the LRCF $C_{xx}(r)$ is only defined on conventional bases, i.e. the ground state $|G\rangle$ (Eq.~\ref{G}) and its Hermitian conjugation $\langle G|$, and all the pair contractions in Eq. ~\ref{AB} are also based on conventional bases of the ground state.

\begin{figure*}[htbp]\centering
\includegraphics[width=13cm]{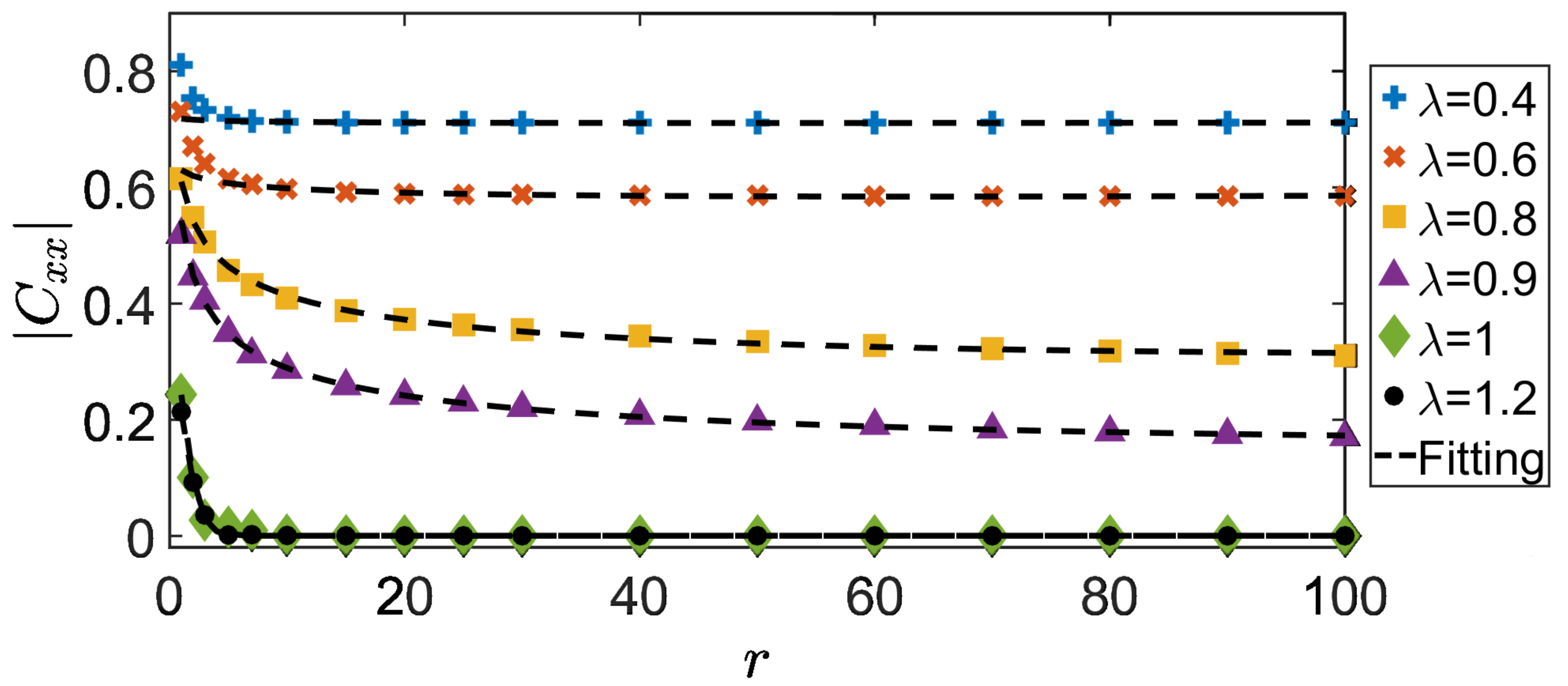}
 \caption{Correlation function $|C_{xx}(r)|$ in different $\lambda$. The other parameters are $(J,\,\gamma,\,\Gamma)=(1,\,0.5,\,1.6)$, corresponding to a gapless point $C_0$ with $\lambda_c=0.6$. Color points come from exact solution. Black dashed lines are obtained by numerical fitting with a function of the form $|C_{xx}(r)|=Ar^{-B}e^{-Cr}$.}\label{Cxx6h}
\end{figure*}

\begin{figure*}[htbp]\centering
\includegraphics[width=17cm]{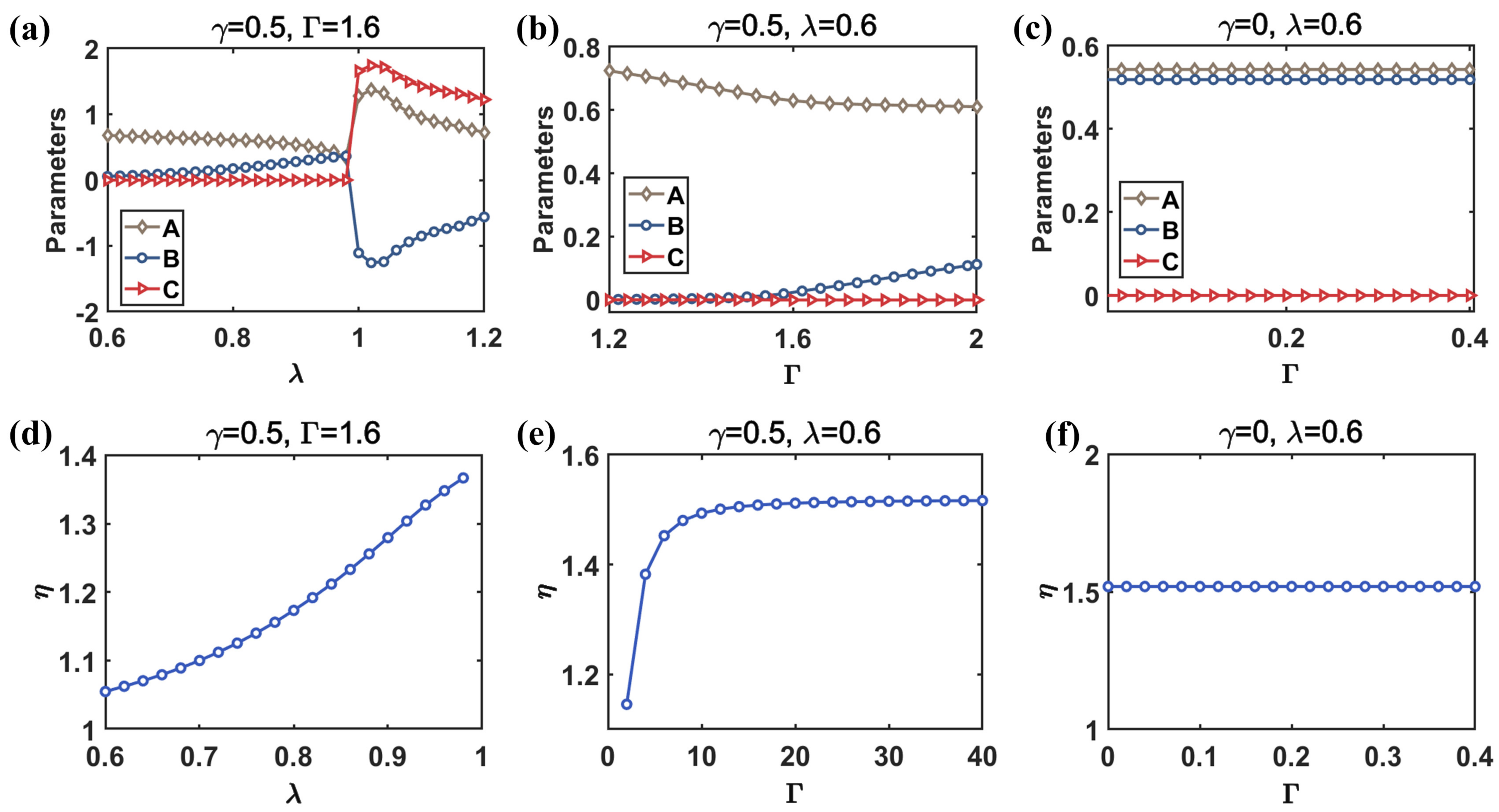}
 \caption{Numerical fitting of LRCF. The fitting function is $|C_{xx}(r)|=Ar^{-B}e^{-Cr}$. (a)-(c) exhibit the fitting parameters $A$ (grey diamond), $B$ (blue circle) and $C$ (red triangle). Panels (d)-(f) are the exponent $\eta$, which equals to $1+B$ and reflects the critical behaviors in CTZ. In (a) the system changes from CTZ to PM phase. In (b) the system changes FM phase to CTZ. The transition point of both (a) and (b) is the EP $C_0 (\lambda=0.6,\, \gamma=0.5,\, \Gamma=1.6)$. In (c) and (f) the system is in CTZ. we set $\gamma=0$ and the system reduces to a non-Hermitian isotropic XY chain. Throughout, J is set as 1.
}\label{ABC}
\end{figure*}

 From Eq.\,(\ref{Cxxr}) to Eq.\,(\ref{AB}), we can obtain the exact value of $C_{xx}(r)$. Here, we show the values of $|C_{xx}(r)|$ in different $\lambda$ with $(J,\gamma,\Gamma)=(1,0.5,1.6)$. The data is plotted in Fig.~\ref{Cxx6h}. When $\lambda=1$ and $\lambda=1.2$, the system is in the region \Rmnum{1} and the LRCFs exponentially decay with the increase of $r$, which implies the system is in PM phase. When $\lambda=0.4$ and $\lambda=0.6$, the system is in the region \Rmnum{2} and the LRCFs are constants when $r$ is large, which shows the characteristic of FM phase. 

 An intriguing result which is unique for non-Hermitian case is obtained for the region \Rmnum{3}. When $\lambda=0.8$ and $\lambda=0.9$, the system is in the region \Rmnum{3}, the decay of $C_{xx}(r)$ is like polynomial type. In the Hermitian system, the polynomial decay only appears at critical point. In the non-Hermitian circumstances, the whole region \Rmnum{3} shows the similar behavior. Therefore, we call region \Rmnum{3} as CTZ. It is worth noting that, in Hermitian system, we can not distinguish ferromagnetic and paramagnetic phases only by energy gap. However, in this model, the different magnetic phases can be characterized by non-Hermitian gap.

To describe the decay more accurately, we fit data points in Fig.~\ref{Cxx6h} by the function
\begin{equation} \label{fitting}
|C_{xx}(r)|=Ar^{-B}e^{-Cr}\;,
\end{equation}
where $A$, $B$, $C$ are the fitting parameters. Obviously, $C>0$ means an exponentially decay of correlation function, while the case $C=0$ and $B>0$ represents the polynomial decay. The fitting curves are shown by the black dashed lines in Fig.~\ref{Cxx6h}.

Now, we study the decay of $C_{xx}(r)$ by scanning the parameters across an exceptional point $C_0 (\lambda=0.6,\, \gamma=0.5,\, \Gamma=1.6)$ along $\lambda$ and $\Gamma$. Fitting by Eq.\,(\ref{fitting}), we get curves of parameters $A$, $B$, $C$ in Fig.~\ref{ABC} (a) and (b). In (a), $\lambda$ changes from $0.6$ to $1.2$. When $\lambda<1$, the system is in CTZ, where the red and blue curves show that $C=0$ and $B>0$. When $\lambda \ge 1$, the system is in region \Rmnum{1}, where the ground state is in PM phase characterized by $C>0$. In (b), we scan $\Gamma$ from $1.2$ to $2$, where $C$ is always equal to zero. When $\Gamma \le 1.6$, the system is in region \Rmnum{2} and $B$ is near zero. Therefore, this region is corresponding to FM phase. In contrast, $B$ is obviously larger than zero when the system is in CTZ with $\Gamma>1.6$.

In CTZ, the parameter $C$ is always equal to zero so that the correlation function $|C_{xx}(r)|$ is proportional to $r^{-B}$. Compared with $C_{xx}(r) \propto r^{1-\eta}$ at critical points in Hermitian XY chain, we have that $\eta=1+B$. Fig.~\ref{ABC}(d) and (e) show $\eta$ in CTZ with $\lambda \in [0.6,0.99]$ and $\Gamma \in [2,40]$, which reflect that non-Hermiticity influences critical behavior. When we change $\lambda$ in CTZ, $\eta$ is no longer standard value ($5/4$) of the Ising transition. When we increase $\Gamma$, $\eta$ tends to be near $3/2$, which corresponds to the anisotropic transition. It may imply that at large $\Gamma$ limit, the critical behavior of the non-Hermitian system is similar to the critical point of anisotropic transition. The intuitive explanation is that when $\Gamma$ is very large, the difference between the interactions in $x$ and $y$ directions, corresponding to $(J+\gamma)/2$ and $(J-\gamma)/2$, becomes comparatively insignificant.

In addition, we point out that the non-Hermitian term $i\frac{\Gamma}{2}\sigma^u_j$ does not influence anisotropic transition. This transition comes from the competition between $\frac{1}{2}(J+\gamma)\sigma^x_j\sigma^x_{j+1}$ and $\frac{1}{2}(J-\gamma)\sigma^y_j\sigma^y_{j+1}$. However, the non-Hermitian term is along $z$ direction ($\sigma^u=\frac{1}{2}(\mathbf{1}+\sigma^z)$), which does not break the symmetry between $x$ and $y$ directions when $\gamma=0$. We show the fitting with $\gamma=0$ in Fig.~\ref{ABC} (c) and (f). It is obvious that $A$, $B$, $C$ and $\eta$ remain unchanged when we increase $\Gamma$. Furthermore, $\eta$ has always been $3/2$, which is same as the value of anisotropic transition in Hermitian XY chain.

At last, we mention the phase diagram again. In Sec.3 we study the derivatives of ground state energy density and find that non-analyticity is only on the EP ring, where the complex gap is equal to 0. This result seems to indicate that the system has only two phases, i.e. inside and outside the EP ring. In fact, by analyzing LRCFs, we find that there are three phases. An extra QPT occurs at the line with $|\lambda/J|=1$, where the gap is a pure imaginary number. This is a new property for non-Hermitian system that the QPT can occur without gap closing. The similar property has been found in the non-Hermitian Kitaev’s toric-code model~\cite{Matsumoto2020}.

\section{discussion and conclusion}
Experimentally, the non-Hermitian XY model can be realized by some artificial quantum systems, such as ultracold atoms in optical lattices, superconducting qubits and coupled cavity arrays. For example, one can use three-level atoms in optical lattices to realize the required Hamiltonian. Two metastable states of atoms represent spin up and spin down. The Hermitian part of the system can be realized by arraying atoms in optical lattices~\cite{Lukin2003,Lukin2013,Zhu2007}. The non-Hermitian term can be generated by exciting one of the metastable states to an auxiliary state~\cite{luole2019}. Besides the cold atom experimental scheme mentioned above, the results can also be realized in coupled cavity arrays~\cite{Jonathan2013}, in which the non Hermiticity can be realized by active and passive cavities~\cite{Xiaomin2014,YangLan2014,Shou2017}. 

In summary, we have investigated the traditional QPT in a quantum XY spin chain with a global complex transverse field. This non-Hermitian transverse field changes the phase diagram of XY model. The results reveal that (i) the second-order QPT points change from Ising transition point to an exceptional ring. (ii) the critical points of Hermitian system are extended to a critical transition zone. In this zone, correlation function decays polynomially. Furthermore, the results reveal the correspondence among the different phases, non-Hermitian gap and LRCF.

Our results indicate the nontrivial influence of non-Hermiticity and our model offers a higher dimensional parameter space to study the critical behaviors and non-Hermitian quantum magnetism. Moreover, our findings in this paper can be readily realized with recent experimental techniques. We believe that our work will benefit the future research on traditional quantum phase transition of non-Hermitian systems.

\section*{Acknowledgements}
We are grateful to S. L. Zhu for dedicated teaching and helpful discussions. This work was supported by the National Natural Science Foundation of China (Grants No. 12074180 and No. 11704132), the Key-Area Research and Development Program of GuangDong Province (Grant No. 2019B030330001), the Key Project of Science and Technology of Guangzhou (Grants No. 201804020055 and No. 2019050001).


\bibliography{bib}

\providecommand{\noopsort}[1]{}\providecommand{\singleletter}[1]{#1}%
\begin{thebibliography}{68}%
\makeatletter
\providecommand \@ifxundefined [1]{%
 \@ifx{#1\undefined}
}%
\providecommand \@ifnum [1]{%
 \ifnum #1\expandafter \@firstoftwo
 \else \expandafter \@secondoftwo
 \fi
}%
\providecommand \@ifx [1]{%
 \ifx #1\expandafter \@firstoftwo
 \else \expandafter \@secondoftwo
 \fi
}%
\providecommand \natexlab [1]{#1}%
\providecommand \enquote  [1]{``#1''}%
\providecommand \bibnamefont  [1]{#1}%
\providecommand \bibfnamefont [1]{#1}%
\providecommand \citenamefont [1]{#1}%
\providecommand \href@noop [0]{\@secondoftwo}%
\providecommand \href [0]{\begingroup \@sanitize@url \@href}%
\providecommand \@href[1]{\@@startlink{#1}\@@href}%
\providecommand \@@href[1]{\endgroup#1\@@endlink}%
\providecommand \@sanitize@url [0]{\catcode `\\12\catcode `\$12\catcode
  `\&12\catcode `\#12\catcode `\^12\catcode `\_12\catcode `\%12\relax}%
\providecommand \@@startlink[1]{}%
\providecommand \@@endlink[0]{}%
\providecommand \url  [0]{\begingroup\@sanitize@url \@url }%
\providecommand \@url [1]{\endgroup\@href {#1}{\urlprefix }}%
\providecommand \urlprefix  [0]{URL }%
\providecommand \Eprint [0]{\href }%
\providecommand \doibase [0]{http://dx.doi.org/}%
\providecommand \selectlanguage [0]{\@gobble}%
\providecommand \bibinfo  [0]{\@secondoftwo}%
\providecommand \bibfield  [0]{\@secondoftwo}%
\providecommand \translation [1]{[#1]}%
\providecommand \BibitemOpen [0]{}%
\providecommand \bibitemStop [0]{}%
\providecommand \bibitemNoStop [0]{.\EOS\space}%
\providecommand \EOS [0]{\spacefactor3000\relax}%
\providecommand \BibitemShut  [1]{\csname bibitem#1\endcsname}%
\let\auto@bib@innerbib\@empty
\bibitem [{\citenamefont {Sachdev}(2011)}]{Sachdev2011}%
  \BibitemOpen
  \bibfield  {author} {\bibinfo {author} {\bibfnamefont {S.}~\bibnamefont
  {Sachdev}},\ }\href {\doibase 10.1017/CBO9780511973765} {\emph {\bibinfo
  {title} {Quantum Phase Transitions (2nd ed.)}}}\ (\bibinfo  {publisher}
  {Cambridge University Press},\ \bibinfo {year} {2011})\BibitemShut {NoStop}%
\bibitem [{\citenamefont {Pfeuty}(1970)}]{Pfeuty1970}%
  \BibitemOpen
  \bibfield  {author} {\bibinfo {author} {\bibfnamefont {P.}~\bibnamefont
  {Pfeuty}},\ }\bibfield  {title} {\emph {\bibinfo {title} {The one-dimensional
  Ising model with a transverse field},\ }}\href {\doibase
  10.1016/0003-4916(70)90270-8} {\bibfield  {journal} {\bibinfo  {journal}
  {Annals of Physics}\ }\textbf {\bibinfo {volume} {57}},\ \bibinfo {pages}
  {79} (\bibinfo {year} {1970})}\BibitemShut {NoStop}%
\bibitem [{\citenamefont {Barouch}\ and\ \citenamefont
  {McCoy}(1971)}]{McCoy1971}%
  \BibitemOpen
  \bibfield  {author} {\bibinfo {author} {\bibfnamefont {E.}~\bibnamefont
  {Barouch}}\ and\ \bibinfo {author} {\bibfnamefont {B.~M.}\ \bibnamefont
  {McCoy}},\ }\bibfield  {title} {\emph {\bibinfo {title} {Statistical
  Mechanics of the $XY$ Model. II. Spin-Correlation Functions},\ }}\href
  {\doibase 10.1103/PhysRevA.3.786} {\bibfield  {journal} {\bibinfo  {journal}
  {Phys. Rev. A}\ }\textbf {\bibinfo {volume} {3}},\ \bibinfo {pages} {786}
  (\bibinfo {year} {1971})}\BibitemShut {NoStop}%
\bibitem [{\citenamefont {Lieb}\ \emph {et~al.}(1961)\citenamefont {Lieb},
  \citenamefont {Schultz},\ and\ \citenamefont {Mattis}}]{Daniel1961}%
  \BibitemOpen
  \bibfield  {author} {\bibinfo {author} {\bibfnamefont {E.}~\bibnamefont
  {Lieb}}, \bibinfo {author} {\bibfnamefont {T.}~\bibnamefont {Schultz}}, \
  and\ \bibinfo {author} {\bibfnamefont {D.}~\bibnamefont {Mattis}},\
  }\bibfield  {title} {\emph {\bibinfo {title} {Two soluble models of an
  antiferromagnetic chain},\ }}\href {\doibase 10.1016/0003-4916(61)90115-4}
  {\bibfield  {journal} {\bibinfo  {journal} {Annals of Physics}\ }\textbf
  {\bibinfo {volume} {16}},\ \bibinfo {pages} {407} (\bibinfo {year}
  {1961})}\BibitemShut {NoStop}%
\bibitem [{\citenamefont {Latorre}\ \emph {et~al.}(2004)\citenamefont
  {Latorre}, \citenamefont {Rico},\ and\ \citenamefont {Vidal}}]{Vidal2004}%
  \BibitemOpen
  \bibfield  {author} {\bibinfo {author} {\bibfnamefont {J.~I.}\ \bibnamefont
  {Latorre}}, \bibinfo {author} {\bibfnamefont {E.}~\bibnamefont {Rico}}, \
  and\ \bibinfo {author} {\bibfnamefont {G.}~\bibnamefont {Vidal}},\ }\bibfield
   {title} {\emph {\bibinfo {title} {Ground state entanglement in quantum spin
  chains},\ }}\href {https://arxiv.org/abs/quant-ph/0304098} {\bibfield
  {journal} {\bibinfo  {journal} {arXiv:quant-ph/0304098}\ } (\bibinfo {year}
  {2004})}\BibitemShut {NoStop}%
\bibitem [{\citenamefont {Jordan}\ and\ \citenamefont
  {Wigner}(1928)}]{Jordan1928}%
  \BibitemOpen
  \bibfield  {author} {\bibinfo {author} {\bibfnamefont {P.}~\bibnamefont
  {Jordan}}\ and\ \bibinfo {author} {\bibfnamefont {E.}~\bibnamefont
  {Wigner}},\ }\bibfield  {title} {\emph {\bibinfo {title} {{\"U}ber das
  Paulische {\"A}quivalenzverbot},\ }}\href {\doibase 10.1007/BF01331938}
  {\bibfield  {journal} {\bibinfo  {journal} {Zeitschrift f{\"u}r Physik}\
  }\textbf {\bibinfo {volume} {47}},\ \bibinfo {pages} {631} (\bibinfo {year}
  {1928})}\BibitemShut {NoStop}%
\bibitem [{\citenamefont {Wilson}(1983)}]{Wilson1983}%
  \BibitemOpen
  \bibfield  {author} {\bibinfo {author} {\bibfnamefont {K.~G.}\ \bibnamefont
  {Wilson}},\ }\bibfield  {title} {\emph {\bibinfo {title} {The renormalization
  group and critical phenomena},\ }}\href {\doibase 10.1103/RevModPhys.55.583}
  {\bibfield  {journal} {\bibinfo  {journal} {Rev. Mod. Phys.}\ }\textbf
  {\bibinfo {volume} {55}},\ \bibinfo {pages} {583} (\bibinfo {year}
  {1983})}\BibitemShut {NoStop}%
\bibitem [{\citenamefont {Jullien}\ \emph {et~al.}(1978)\citenamefont
  {Jullien}, \citenamefont {Pfeuty}, \citenamefont {Fields},\ and\
  \citenamefont {Doniach}}]{Doniach1978}%
  \BibitemOpen
  \bibfield  {author} {\bibinfo {author} {\bibfnamefont {R.}~\bibnamefont
  {Jullien}}, \bibinfo {author} {\bibfnamefont {P.}~\bibnamefont {Pfeuty}},
  \bibinfo {author} {\bibfnamefont {J.~N.}\ \bibnamefont {Fields}}, \ and\
  \bibinfo {author} {\bibfnamefont {S.}~\bibnamefont {Doniach}},\ }\bibfield
  {title} {\emph {\bibinfo {title} {Zero-temperature renormalization method for
  quantum systems. I. Ising model in a transverse field in one dimension},\
  }}\href {\doibase 10.1103/PhysRevB.18.3568} {\bibfield  {journal} {\bibinfo
  {journal} {Phys. Rev. B}\ }\textbf {\bibinfo {volume} {18}},\ \bibinfo
  {pages} {3568} (\bibinfo {year} {1978})}\BibitemShut {NoStop}%
\bibitem [{\citenamefont {Jullien}\ and\ \citenamefont
  {Pfeuty}(1979)}]{Pfeuty1979}%
  \BibitemOpen
  \bibfield  {author} {\bibinfo {author} {\bibfnamefont {R.}~\bibnamefont
  {Jullien}}\ and\ \bibinfo {author} {\bibfnamefont {P.}~\bibnamefont
  {Pfeuty}},\ }\bibfield  {title} {\emph {\bibinfo {title} {Zero-temperature
  renormalization-group method for quantum systems. II. Isotropic
  $X\ensuremath{-}Y$ model in a transverse field in one dimension},\ }}\href
  {\doibase 10.1103/PhysRevB.19.4646} {\bibfield  {journal} {\bibinfo
  {journal} {Phys. Rev. B}\ }\textbf {\bibinfo {volume} {19}},\ \bibinfo
  {pages} {4646} (\bibinfo {year} {1979})}\BibitemShut {NoStop}%
\bibitem [{\citenamefont {Kargarian}\ \emph {et~al.}(2008)\citenamefont
  {Kargarian}, \citenamefont {Jafari},\ and\ \citenamefont
  {Langari}}]{Langari2008}%
  \BibitemOpen
  \bibfield  {author} {\bibinfo {author} {\bibfnamefont {M.}~\bibnamefont
  {Kargarian}}, \bibinfo {author} {\bibfnamefont {R.}~\bibnamefont {Jafari}}, \
  and\ \bibinfo {author} {\bibfnamefont {A.}~\bibnamefont {Langari}},\
  }\bibfield  {title} {\emph {\bibinfo {title} {Renormalization of entanglement
  in the anisotropic Heisenberg $(XXZ)$ model},\ }}\href {\doibase
  10.1103/PhysRevA.77.032346} {\bibfield  {journal} {\bibinfo  {journal} {Phys.
  Rev. A}\ }\textbf {\bibinfo {volume} {77}},\ \bibinfo {pages} {032346}
  (\bibinfo {year} {2008})}\BibitemShut {NoStop}%
\bibitem [{\citenamefont {Tong}\ and\ \citenamefont {Zhong}(2001)}]{Ming2001}%
  \BibitemOpen
  \bibfield  {author} {\bibinfo {author} {\bibfnamefont {P.}~\bibnamefont
  {Tong}}\ and\ \bibinfo {author} {\bibfnamefont {M.}~\bibnamefont {Zhong}},\
  }\bibfield  {title} {\emph {\bibinfo {title} {Quantum phase transitions of
  periodic anisotropic XY chain in a transverse field},\ }}\href {\doibase
  10.1016/S0921-4526(01)00546-4} {\bibfield  {journal} {\bibinfo  {journal}
  {Physica B: Condensed Matter}\ }\textbf {\bibinfo {volume} {304}},\ \bibinfo
  {pages} {91} (\bibinfo {year} {2001})}\BibitemShut {NoStop}%
\bibitem [{\citenamefont {Mofidnakhaei}\ \emph {et~al.}(2018)\citenamefont
  {Mofidnakhaei}, \citenamefont {Fumani}, \citenamefont {Mahdavifar},\ and\
  \citenamefont {Vahedi}}]{Vahedi2018}%
  \BibitemOpen
  \bibfield  {author} {\bibinfo {author} {\bibfnamefont {F.}~\bibnamefont
  {Mofidnakhaei}}, \bibinfo {author} {\bibfnamefont {F.}~\bibnamefont
  {Fumani}}, \bibinfo {author} {\bibfnamefont {S.}~\bibnamefont {Mahdavifar}},
  \ and\ \bibinfo {author} {\bibfnamefont {J.}~\bibnamefont {Vahedi}},\
  }\bibfield  {title} {\emph {\bibinfo {title} {Quantum correlations in
  anisotropic XY-spin chains in a transverse magnetic field},\ }}\href
  {\doibase 10.1080/01411594.2018.1527916} {\bibfield  {journal} {\bibinfo
  {journal} {Phase Transitions}\ }\textbf {\bibinfo {volume} {91}},\ \bibinfo
  {pages} {1256} (\bibinfo {year} {2018})}\BibitemShut {NoStop}%
\bibitem [{\citenamefont {Zhang}\ \emph {et~al.}(2017)\citenamefont {Zhang},
  \citenamefont {Li},\ and\ \citenamefont {Song}}]{Song2017}%
  \BibitemOpen
  \bibfield  {author} {\bibinfo {author} {\bibfnamefont {G.}~\bibnamefont
  {Zhang}}, \bibinfo {author} {\bibfnamefont {C.}~\bibnamefont {Li}}, \ and\
  \bibinfo {author} {\bibfnamefont {Z.}~\bibnamefont {Song}},\ }\bibfield
  {title} {\emph {\bibinfo {title} {Majorana charges, winding numbers and Chern
  numbers in quantum Ising models},\ }}\href {\doibase
  10.1038/s41598-017-08323-0} {\bibfield  {journal} {\bibinfo  {journal}
  {Scientific Reports}\ }\textbf {\bibinfo {volume} {7}},\ \bibinfo {pages}
  {8176} (\bibinfo {year} {2017})}\BibitemShut {NoStop}%
\bibitem [{\citenamefont {Zhang}\ and\ \citenamefont {Song}(2015)}]{Song2015}%
  \BibitemOpen
  \bibfield  {author} {\bibinfo {author} {\bibfnamefont {G.}~\bibnamefont
  {Zhang}}\ and\ \bibinfo {author} {\bibfnamefont {Z.}~\bibnamefont {Song}},\
  }\bibfield  {title} {\emph {\bibinfo {title} {Topological Characterization of
  Extended Quantum Ising Models},\ }}\href {\doibase
  10.1103/PhysRevLett.115.177204} {\bibfield  {journal} {\bibinfo  {journal}
  {Phys. Rev. Lett.}\ }\textbf {\bibinfo {volume} {115}},\ \bibinfo {pages}
  {177204} (\bibinfo {year} {2015})}\BibitemShut {NoStop}%
\bibitem [{\citenamefont {Bunder}\ and\ \citenamefont
  {McKenzie}(1999)}]{Ross1999}%
  \BibitemOpen
  \bibfield  {author} {\bibinfo {author} {\bibfnamefont {J.~E.}\ \bibnamefont
  {Bunder}}\ and\ \bibinfo {author} {\bibfnamefont {R.~H.}\ \bibnamefont
  {McKenzie}},\ }\bibfield  {title} {\emph {\bibinfo {title} {Effect of
  disorder on quantum phase transitions in anisotropic XY spin chains in a
  transverse field},\ }}\href {\doibase 10.1103/PhysRevB.60.344} {\bibfield
  {journal} {\bibinfo  {journal} {Phys. Rev. B}\ }\textbf {\bibinfo {volume}
  {60}},\ \bibinfo {pages} {344} (\bibinfo {year} {1999})}\BibitemShut
  {NoStop}%
\bibitem [{\citenamefont {Zhong}\ and\ \citenamefont {Tong}(2010)}]{Tong2010}%
  \BibitemOpen
  \bibfield  {author} {\bibinfo {author} {\bibfnamefont {M.}~\bibnamefont
  {Zhong}}\ and\ \bibinfo {author} {\bibfnamefont {P.}~\bibnamefont {Tong}},\
  }\bibfield  {title} {\emph {\bibinfo {title} {The Ising and anisotropy phase
  transitions of the periodic XY model in a transverse field},\ }}\href
  {\doibase 10.1088/1751-8113/43/50/505302} {\bibfield  {journal} {\bibinfo
  {journal} {Journal of Physics A: Mathematical and Theoretical}\ }\textbf
  {\bibinfo {volume} {43}},\ \bibinfo {pages} {505302} (\bibinfo {year}
  {2010})}\BibitemShut {NoStop}%
\bibitem [{\citenamefont {Zhong}\ \emph {et~al.}(2013)\citenamefont {Zhong},
  \citenamefont {Xu}, \citenamefont {Liu},\ and\ \citenamefont
  {Tong}}]{Tong2013}%
  \BibitemOpen
  \bibfield  {author} {\bibinfo {author} {\bibfnamefont {M.}~\bibnamefont
  {Zhong}}, \bibinfo {author} {\bibfnamefont {H.}~\bibnamefont {Xu}}, \bibinfo
  {author} {\bibfnamefont {X.-X.}\ \bibnamefont {Liu}}, \ and\ \bibinfo
  {author} {\bibfnamefont {P.-Q.}\ \bibnamefont {Tong}},\ }\bibfield  {title}
  {\emph {\bibinfo {title} {The effects of the
  Dzyaloshinskii{\textemdash}Moriya interaction on the ground-state properties
  of {theXYchain} in a transverse field},\ }}\href {\doibase
  10.1088/1674-1056/22/9/090313} {\bibfield  {journal} {\bibinfo  {journal}
  {Chinese Physics B}\ }\textbf {\bibinfo {volume} {22}},\ \bibinfo {pages}
  {090313} (\bibinfo {year} {2013})}\BibitemShut {NoStop}%
\bibitem [{\citenamefont {Marzolino}\ and\ \citenamefont
  {Prosen}(2017)}]{Ugo2017}%
  \BibitemOpen
  \bibfield  {author} {\bibinfo {author} {\bibfnamefont {U.}~\bibnamefont
  {Marzolino}}\ and\ \bibinfo {author} {\bibfnamefont {T.~c.~v.}\ \bibnamefont
  {Prosen}},\ }\bibfield  {title} {\emph {\bibinfo {title} {Fisher information
  approach to nonequilibrium phase transitions in a quantum XXZ spin chain with
  boundary noise},\ }}\href {\doibase 10.1103/PhysRevB.96.104402} {\bibfield
  {journal} {\bibinfo  {journal} {Phys. Rev. B}\ }\textbf {\bibinfo {volume}
  {96}},\ \bibinfo {pages} {104402} (\bibinfo {year} {2017})}\BibitemShut
  {NoStop}%
\bibitem [{\citenamefont {Gao}\ \emph {et~al.}(2017)\citenamefont {Gao},
  \citenamefont {Zhang}, \citenamefont {Yu},\ and\ \citenamefont
  {Zhu}}]{Gao2017}%
  \BibitemOpen
  \bibfield  {author} {\bibinfo {author} {\bibfnamefont {Z.-P.}\ \bibnamefont
  {Gao}}, \bibinfo {author} {\bibfnamefont {D.-W.}\ \bibnamefont {Zhang}},
  \bibinfo {author} {\bibfnamefont {Y.}~\bibnamefont {Yu}}, \ and\ \bibinfo
  {author} {\bibfnamefont {S.-L.}\ \bibnamefont {Zhu}},\ }\bibfield  {title}
  {\emph {\bibinfo {title} {Anti-Kibble-Zurek behavior of a noisy
  transverse-field $\mathrm{XY}$ chain and its quantum simulation with
  two-level systems},\ }}\href {\doibase 10.1103/PhysRevB.95.224303} {\bibfield
   {journal} {\bibinfo  {journal} {Phys. Rev. B}\ }\textbf {\bibinfo {volume}
  {95}},\ \bibinfo {pages} {224303} (\bibinfo {year} {2017})}\BibitemShut
  {NoStop}%
\bibitem [{\citenamefont {Zhu}(2006)}]{Zhu2006}%
  \BibitemOpen
  \bibfield  {author} {\bibinfo {author} {\bibfnamefont {S.-L.}\ \bibnamefont
  {Zhu}},\ }\bibfield  {title} {\emph {\bibinfo {title} {Scaling of Geometric
  Phases Close to the Quantum Phase Transition in the $XY$ Spin Chain},\
  }}\href {\doibase 10.1103/PhysRevLett.96.077206} {\bibfield  {journal}
  {\bibinfo  {journal} {Phys. Rev. Lett.}\ }\textbf {\bibinfo {volume} {96}},\
  \bibinfo {pages} {077206} (\bibinfo {year} {2006})}\BibitemShut {NoStop}%
\bibitem [{\citenamefont {Bi}\ \emph {et~al.}(2021)\citenamefont {Bi},
  \citenamefont {He},\ and\ \citenamefont {Li}}]{Lipeng2020}%
  \BibitemOpen
  \bibfield  {author} {\bibinfo {author} {\bibfnamefont {S.}~\bibnamefont
  {Bi}}, \bibinfo {author} {\bibfnamefont {Y.}~\bibnamefont {He}}, \ and\
  \bibinfo {author} {\bibfnamefont {P.}~\bibnamefont {Li}},\ }\bibfield
  {title} {\emph {\bibinfo {title} {Ring-frustrated non-Hermitian XY model},\
  }}\href {\doibase 10.1016/j.physleta.2021.127208} {\bibfield  {journal}
  {\bibinfo  {journal} {Physics Letters A}\ }\textbf {\bibinfo {volume}
  {395}},\ \bibinfo {pages} {127208} (\bibinfo {year} {2021})}\BibitemShut
  {NoStop}%
\bibitem [{\citenamefont {Li}\ \emph {et~al.}(2019)\citenamefont {Li},
  \citenamefont {Harter}, \citenamefont {Liu}, \citenamefont {de~Melo},
  \citenamefont {Joglekar},\ and\ \citenamefont {Luo}}]{luole2019}%
  \BibitemOpen
  \bibfield  {author} {\bibinfo {author} {\bibfnamefont {J.}~\bibnamefont
  {Li}}, \bibinfo {author} {\bibfnamefont {A.~K.}\ \bibnamefont {Harter}},
  \bibinfo {author} {\bibfnamefont {J.}~\bibnamefont {Liu}}, \bibinfo {author}
  {\bibfnamefont {L.}~\bibnamefont {de~Melo}}, \bibinfo {author} {\bibfnamefont
  {Y.~N.}\ \bibnamefont {Joglekar}}, \ and\ \bibinfo {author} {\bibfnamefont
  {L.}~\bibnamefont {Luo}},\ }\bibfield  {title} {\emph {\bibinfo {title}
  {Observation of parity-time symmetry breaking transitions in a dissipative
  Floquet system of ultracold atoms},\ }}\href {\doibase
  10.1038/s41467-019-08596-1} {\bibfield  {journal} {\bibinfo  {journal}
  {Nature Communications}\ }\textbf {\bibinfo {volume} {10}},\ \bibinfo {pages}
  {855} (\bibinfo {year} {2019})}\BibitemShut {NoStop}%
\bibitem [{\citenamefont {Li}\ \emph {et~al.}(2020)\citenamefont {Li},
  \citenamefont {Lee},\ and\ \citenamefont {Gong}}]{Jiangbin2020}%
  \BibitemOpen
  \bibfield  {author} {\bibinfo {author} {\bibfnamefont {L.}~\bibnamefont
  {Li}}, \bibinfo {author} {\bibfnamefont {C.~H.}\ \bibnamefont {Lee}}, \ and\
  \bibinfo {author} {\bibfnamefont {J.}~\bibnamefont {Gong}},\ }\bibfield
  {title} {\emph {\bibinfo {title} {Topological Switch for Non-Hermitian Skin
  Effect in Cold-Atom Systems with Loss},\ }}\href {\doibase
  10.1103/PhysRevLett.124.250402} {\bibfield  {journal} {\bibinfo  {journal}
  {Phys. Rev. Lett.}\ }\textbf {\bibinfo {volume} {124}},\ \bibinfo {pages}
  {250402} (\bibinfo {year} {2020})}\BibitemShut {NoStop}%
\bibitem [{\citenamefont {Chen}\ \emph {et~al.}(2017)\citenamefont {Chen},
  \citenamefont {Kaya~{\"O}zdemir}, \citenamefont {Zhao}, \citenamefont
  {Wiersig},\ and\ \citenamefont {Yang}}]{YangLan2017}%
  \BibitemOpen
  \bibfield  {author} {\bibinfo {author} {\bibfnamefont {W.}~\bibnamefont
  {Chen}}, \bibinfo {author} {\bibfnamefont {{\c S}.}~\bibnamefont
  {Kaya~{\"O}zdemir}}, \bibinfo {author} {\bibfnamefont {G.}~\bibnamefont
  {Zhao}}, \bibinfo {author} {\bibfnamefont {J.}~\bibnamefont {Wiersig}}, \
  and\ \bibinfo {author} {\bibfnamefont {L.}~\bibnamefont {Yang}},\ }\bibfield
  {title} {\emph {\bibinfo {title} {Exceptional points enhance sensing in an
  optical microcavity},\ }}\href {\doibase 10.1038/nature23281} {\bibfield
  {journal} {\bibinfo  {journal} {Nature}\ }\textbf {\bibinfo {volume} {548}},\
  \bibinfo {pages} {192} (\bibinfo {year} {2017})}\BibitemShut {NoStop}%
\bibitem [{\citenamefont {Peng}\ \emph {et~al.}(2014)\citenamefont {Peng},
  \citenamefont {{\"O}zdemir}, \citenamefont {Lei}, \citenamefont {Monifi},
  \citenamefont {Gianfreda}, \citenamefont {Long}, \citenamefont {Fan},
  \citenamefont {Nori}, \citenamefont {Bender},\ and\ \citenamefont
  {Yang}}]{YangLan2014}%
  \BibitemOpen
  \bibfield  {author} {\bibinfo {author} {\bibfnamefont {B.}~\bibnamefont
  {Peng}}, \bibinfo {author} {\bibfnamefont {{\c S}.~K.}\ \bibnamefont
  {{\"O}zdemir}}, \bibinfo {author} {\bibfnamefont {F.}~\bibnamefont {Lei}},
  \bibinfo {author} {\bibfnamefont {F.}~\bibnamefont {Monifi}}, \bibinfo
  {author} {\bibfnamefont {M.}~\bibnamefont {Gianfreda}}, \bibinfo {author}
  {\bibfnamefont {G.~L.}\ \bibnamefont {Long}}, \bibinfo {author}
  {\bibfnamefont {S.}~\bibnamefont {Fan}}, \bibinfo {author} {\bibfnamefont
  {F.}~\bibnamefont {Nori}}, \bibinfo {author} {\bibfnamefont {C.~M.}\
  \bibnamefont {Bender}}, \ and\ \bibinfo {author} {\bibfnamefont
  {L.}~\bibnamefont {Yang}},\ }\bibfield  {title} {\emph {\bibinfo {title}
  {Parity-time-symmetric whispering-gallery microcavities},\ }}\href {\doibase
  10.1038/nphys2927} {\bibfield  {journal} {\bibinfo  {journal} {Nature
  Physics}\ }\textbf {\bibinfo {volume} {10}},\ \bibinfo {pages} {394}
  (\bibinfo {year} {2014})}\BibitemShut {NoStop}%
\bibitem [{\citenamefont {Cerjan}\ \emph {et~al.}(2019)\citenamefont {Cerjan},
  \citenamefont {Huang}, \citenamefont {Wang}, \citenamefont {Chen},
  \citenamefont {Chong},\ and\ \citenamefont {Rechtsman}}]{Rechtsman2019}%
  \BibitemOpen
  \bibfield  {author} {\bibinfo {author} {\bibfnamefont {A.}~\bibnamefont
  {Cerjan}}, \bibinfo {author} {\bibfnamefont {S.}~\bibnamefont {Huang}},
  \bibinfo {author} {\bibfnamefont {M.}~\bibnamefont {Wang}}, \bibinfo {author}
  {\bibfnamefont {K.~P.}\ \bibnamefont {Chen}}, \bibinfo {author}
  {\bibfnamefont {Y.}~\bibnamefont {Chong}}, \ and\ \bibinfo {author}
  {\bibfnamefont {M.~C.}\ \bibnamefont {Rechtsman}},\ }\bibfield  {title}
  {\emph {\bibinfo {title} {Experimental realization of a Weyl exceptional
  ring},\ }}\href {\doibase 10.1038/s41566-019-0453-z} {\bibfield  {journal}
  {\bibinfo  {journal} {Nature Photonics}\ }\textbf {\bibinfo {volume} {13}},\
  \bibinfo {pages} {623} (\bibinfo {year} {2019})}\BibitemShut {NoStop}%
\bibitem [{\citenamefont {Wu}\ \emph {et~al.}(2019)\citenamefont {Wu},
  \citenamefont {Liu}, \citenamefont {Geng}, \citenamefont {Song},
  \citenamefont {Ye}, \citenamefont {Duan}, \citenamefont {Rong},\ and\
  \citenamefont {Du}}]{Jiangfeng2019}%
  \BibitemOpen
  \bibfield  {author} {\bibinfo {author} {\bibfnamefont {Y.}~\bibnamefont
  {Wu}}, \bibinfo {author} {\bibfnamefont {W.}~\bibnamefont {Liu}}, \bibinfo
  {author} {\bibfnamefont {J.}~\bibnamefont {Geng}}, \bibinfo {author}
  {\bibfnamefont {X.}~\bibnamefont {Song}}, \bibinfo {author} {\bibfnamefont
  {X.}~\bibnamefont {Ye}}, \bibinfo {author} {\bibfnamefont {C.-K.}\
  \bibnamefont {Duan}}, \bibinfo {author} {\bibfnamefont {X.}~\bibnamefont
  {Rong}}, \ and\ \bibinfo {author} {\bibfnamefont {J.}~\bibnamefont {Du}},\
  }\bibfield  {title} {\emph {\bibinfo {title} {Observation of parity-time
  symmetry breaking in a single-spin system},\ }}\href {\doibase
  10.1126/science.aaw8205} {\bibfield  {journal} {\bibinfo  {journal}
  {Science}\ }\textbf {\bibinfo {volume} {364}},\ \bibinfo {pages} {878}
  (\bibinfo {year} {2019})}\BibitemShut {NoStop}%
\bibitem [{\citenamefont {Bender}\ and\ \citenamefont
  {Boettcher}(1998)}]{Bender1998}%
  \BibitemOpen
  \bibfield  {author} {\bibinfo {author} {\bibfnamefont {C.~M.}\ \bibnamefont
  {Bender}}\ and\ \bibinfo {author} {\bibfnamefont {S.}~\bibnamefont
  {Boettcher}},\ }\bibfield  {title} {\emph {\bibinfo {title} {Real Spectra in
  Non-Hermitian Hamiltonians Having $\mathcal{P}$ $\mathcal{T}$ Symmetry},\
  }}\href {\doibase 10.1103/PhysRevLett.80.5243} {\bibfield  {journal}
  {\bibinfo  {journal} {Phys. Rev. Lett.}\ }\textbf {\bibinfo {volume} {80}},\
  \bibinfo {pages} {5243} (\bibinfo {year} {1998})}\BibitemShut {NoStop}%
\bibitem [{\citenamefont {Yao}\ and\ \citenamefont
  {Wang}(2018)}]{YaoShunyu2018}%
  \BibitemOpen
  \bibfield  {author} {\bibinfo {author} {\bibfnamefont {S.}~\bibnamefont
  {Yao}}\ and\ \bibinfo {author} {\bibfnamefont {Z.}~\bibnamefont {Wang}},\
  }\bibfield  {title} {\emph {\bibinfo {title} {Edge States and Topological
  Invariants of Non-Hermitian Systems},\ }}\href {\doibase
  10.1103/PhysRevLett.121.086803} {\bibfield  {journal} {\bibinfo  {journal}
  {Phys. Rev. Lett.}\ }\textbf {\bibinfo {volume} {121}},\ \bibinfo {pages}
  {086803} (\bibinfo {year} {2018})}\BibitemShut {NoStop}%
\bibitem [{\citenamefont {Yao}\ \emph {et~al.}(2018)\citenamefont {Yao},
  \citenamefont {Song},\ and\ \citenamefont {Wang}}]{WangZhong2018}%
  \BibitemOpen
  \bibfield  {author} {\bibinfo {author} {\bibfnamefont {S.}~\bibnamefont
  {Yao}}, \bibinfo {author} {\bibfnamefont {F.}~\bibnamefont {Song}}, \ and\
  \bibinfo {author} {\bibfnamefont {Z.}~\bibnamefont {Wang}},\ }\bibfield
  {title} {\emph {\bibinfo {title} {Non-Hermitian Chern Bands},\ }}\href
  {\doibase 10.1103/PhysRevLett.121.136802} {\bibfield  {journal} {\bibinfo
  {journal} {Phys. Rev. Lett.}\ }\textbf {\bibinfo {volume} {121}},\ \bibinfo
  {pages} {136802} (\bibinfo {year} {2018})}\BibitemShut {NoStop}%
\bibitem [{\citenamefont {Zhang}\ \emph
  {et~al.}(2020{\natexlab{a}})\citenamefont {Zhang}, \citenamefont {Chen},
  \citenamefont {Zhang}, \citenamefont {Lang}, \citenamefont {Li},\ and\
  \citenamefont {Zhu}}]{Danwei2020}%
  \BibitemOpen
  \bibfield  {author} {\bibinfo {author} {\bibfnamefont {D.-W.}\ \bibnamefont
  {Zhang}}, \bibinfo {author} {\bibfnamefont {Y.-L.}\ \bibnamefont {Chen}},
  \bibinfo {author} {\bibfnamefont {G.-Q.}\ \bibnamefont {Zhang}}, \bibinfo
  {author} {\bibfnamefont {L.-J.}\ \bibnamefont {Lang}}, \bibinfo {author}
  {\bibfnamefont {Z.}~\bibnamefont {Li}}, \ and\ \bibinfo {author}
  {\bibfnamefont {S.-L.}\ \bibnamefont {Zhu}},\ }\bibfield  {title} {\emph
  {\bibinfo {title} {Skin superfluid, topological Mott insulators, and
  asymmetric dynamics in an interacting non-Hermitian Aubry-Andr\'e-Harper
  model},\ }}\href {\doibase 10.1103/PhysRevB.101.235150} {\bibfield  {journal}
  {\bibinfo  {journal} {Phys. Rev. B}\ }\textbf {\bibinfo {volume} {101}},\
  \bibinfo {pages} {235150} (\bibinfo {year} {2020}{\natexlab{a}})}\BibitemShut
  {NoStop}%
\bibitem [{\citenamefont {He}\ \emph {et~al.}(2020)\citenamefont {He},
  \citenamefont {Fu}, \citenamefont {Zhang},\ and\ \citenamefont
  {Zhu}}]{Hepeng2020}%
  \BibitemOpen
  \bibfield  {author} {\bibinfo {author} {\bibfnamefont {P.}~\bibnamefont
  {He}}, \bibinfo {author} {\bibfnamefont {J.-H.}\ \bibnamefont {Fu}}, \bibinfo
  {author} {\bibfnamefont {D.-W.}\ \bibnamefont {Zhang}}, \ and\ \bibinfo
  {author} {\bibfnamefont {S.-L.}\ \bibnamefont {Zhu}},\ }\bibfield  {title}
  {\emph {\bibinfo {title} {Double exceptional links in a three-dimensional
  dissipative cold atomic gas},\ }}\href {\doibase 10.1103/PhysRevA.102.023308}
  {\bibfield  {journal} {\bibinfo  {journal} {Phys. Rev. A}\ }\textbf {\bibinfo
  {volume} {102}},\ \bibinfo {pages} {023308} (\bibinfo {year}
  {2020})}\BibitemShut {NoStop}%
\bibitem [{\citenamefont {Shen}\ \emph {et~al.}(2018)\citenamefont {Shen},
  \citenamefont {Zhen},\ and\ \citenamefont {Fu}}]{Fuliang2018}%
  \BibitemOpen
  \bibfield  {author} {\bibinfo {author} {\bibfnamefont {H.}~\bibnamefont
  {Shen}}, \bibinfo {author} {\bibfnamefont {B.}~\bibnamefont {Zhen}}, \ and\
  \bibinfo {author} {\bibfnamefont {L.}~\bibnamefont {Fu}},\ }\bibfield
  {title} {\emph {\bibinfo {title} {Topological Band Theory for Non-Hermitian
  Hamiltonians},\ }}\href {\doibase 10.1103/PhysRevLett.120.146402} {\bibfield
  {journal} {\bibinfo  {journal} {Phys. Rev. Lett.}\ }\textbf {\bibinfo
  {volume} {120}},\ \bibinfo {pages} {146402} (\bibinfo {year}
  {2018})}\BibitemShut {NoStop}%
\bibitem [{\citenamefont {Bergholtz}\ \emph {et~al.}(2021)\citenamefont
  {Bergholtz}, \citenamefont {Budich},\ and\ \citenamefont
  {Kunst}}]{Flore2019}%
  \BibitemOpen
  \bibfield  {author} {\bibinfo {author} {\bibfnamefont {E.~J.}\ \bibnamefont
  {Bergholtz}}, \bibinfo {author} {\bibfnamefont {J.~C.}\ \bibnamefont
  {Budich}}, \ and\ \bibinfo {author} {\bibfnamefont {F.~K.}\ \bibnamefont
  {Kunst}},\ }\bibfield  {title} {\emph {\bibinfo {title} {Exceptional topology
  of non-Hermitian systems},\ }}\href {\doibase 10.1103/RevModPhys.93.015005}
  {\bibfield  {journal} {\bibinfo  {journal} {Rev. Mod. Phys.}\ }\textbf
  {\bibinfo {volume} {93}},\ \bibinfo {pages} {015005} (\bibinfo {year}
  {2021})}\BibitemShut {NoStop}%
\bibitem [{\citenamefont {Ghatak}\ and\ \citenamefont
  {Das}(2019)}]{Ghatak2019}%
  \BibitemOpen
  \bibfield  {author} {\bibinfo {author} {\bibfnamefont {A.}~\bibnamefont
  {Ghatak}}\ and\ \bibinfo {author} {\bibfnamefont {T.}~\bibnamefont {Das}},\
  }\bibfield  {title} {\emph {\bibinfo {title} {New topological invariants in
  non-Hermitian systems},\ }}\href {\doibase 10.1088/1361-648x/ab11b3}
  {\bibfield  {journal} {\bibinfo  {journal} {Journal of Physics: Condensed
  Matter}\ }\textbf {\bibinfo {volume} {31}},\ \bibinfo {pages} {263001}
  (\bibinfo {year} {2019})}\BibitemShut {NoStop}%
\bibitem [{\citenamefont {Kawabata}\ \emph {et~al.}(2019)\citenamefont
  {Kawabata}, \citenamefont {Shiozaki}, \citenamefont {Ueda},\ and\
  \citenamefont {Sato}}]{Masatoshi2019}%
  \BibitemOpen
  \bibfield  {author} {\bibinfo {author} {\bibfnamefont {K.}~\bibnamefont
  {Kawabata}}, \bibinfo {author} {\bibfnamefont {K.}~\bibnamefont {Shiozaki}},
  \bibinfo {author} {\bibfnamefont {M.}~\bibnamefont {Ueda}}, \ and\ \bibinfo
  {author} {\bibfnamefont {M.}~\bibnamefont {Sato}},\ }\bibfield  {title}
  {\emph {\bibinfo {title} {Symmetry and Topology in Non-Hermitian Physics},\
  }}\href {\doibase 10.1103/PhysRevX.9.041015} {\bibfield  {journal} {\bibinfo
  {journal} {Phys. Rev. X}\ }\textbf {\bibinfo {volume} {9}},\ \bibinfo {pages}
  {041015} (\bibinfo {year} {2019})}\BibitemShut {NoStop}%
\bibitem [{\citenamefont {He}\ and\ \citenamefont {Chien}(2020)}]{Heyan2020}%
  \BibitemOpen
  \bibfield  {author} {\bibinfo {author} {\bibfnamefont {Y.}~\bibnamefont
  {He}}\ and\ \bibinfo {author} {\bibfnamefont {C.-C.}\ \bibnamefont {Chien}},\
  }\bibfield  {title} {\emph {\bibinfo {title} {Non-Hermitian generalizations
  of extended Su-Schrieffer-Heeger models},\ }}\href
  {https://iopscience.iop.org/article/10.1088/1361-648X/abc974} {\bibfield
  {journal} {\bibinfo  {journal} {Journal of Physics: Condensed Matter}\
  }\textbf {\bibinfo {volume} {33}} (\bibinfo {year} {2020})}\BibitemShut
  {NoStop}%
\bibitem [{\citenamefont {Zhang}\ \emph
  {et~al.}(2020{\natexlab{b}})\citenamefont {Zhang}, \citenamefont {Tang},
  \citenamefont {Lang}, \citenamefont {Yan},\ and\ \citenamefont
  {Zhu}}]{Zhang2020}%
  \BibitemOpen
  \bibfield  {author} {\bibinfo {author} {\bibfnamefont {D.-W.}\ \bibnamefont
  {Zhang}}, \bibinfo {author} {\bibfnamefont {L.-Z.}\ \bibnamefont {Tang}},
  \bibinfo {author} {\bibfnamefont {L.-J.}\ \bibnamefont {Lang}}, \bibinfo
  {author} {\bibfnamefont {h.}~\bibnamefont {Yan}}, \ and\ \bibinfo {author}
  {\bibfnamefont {S.-L.}\ \bibnamefont {Zhu}},\ }\bibfield  {title} {\emph
  {\bibinfo {title} {Non-hermitian topological anderson insulators},\ }}\href
  {\doibase 10.1007/s11433-020-1521-9} {\bibfield  {journal} {\bibinfo
  {journal} {Science China Physics, Mechanics and Astronomy}\ }\textbf
  {\bibinfo {volume} {63}},\ \bibinfo {pages} {267062} (\bibinfo {year}
  {2020}{\natexlab{b}})}\BibitemShut {NoStop}%
\bibitem [{\citenamefont {Luo}\ and\ \citenamefont
  {Zhang}(2019)}]{Chuanwei2020}%
  \BibitemOpen
  \bibfield  {author} {\bibinfo {author} {\bibfnamefont {X.-W.}\ \bibnamefont
  {Luo}}\ and\ \bibinfo {author} {\bibfnamefont {C.}~\bibnamefont {Zhang}},\
  }\bibfield  {title} {\emph {\bibinfo {title} {Non-Hermitian Disorder-induced
  Topological insulators},\ }}\href {https://arxiv.org/abs/1912.10652}
  {\bibfield  {journal} {\bibinfo  {journal} {arXiv:1912.10652
  [cond-mat.mes-hall]}\ } (\bibinfo {year} {2019})}\BibitemShut {NoStop}%
\bibitem [{\citenamefont {Tang}\ \emph {et~al.}(2020)\citenamefont {Tang},
  \citenamefont {Zhang}, \citenamefont {Zhang},\ and\ \citenamefont
  {Zhang}}]{Tang2020}%
  \BibitemOpen
  \bibfield  {author} {\bibinfo {author} {\bibfnamefont {L.-Z.}\ \bibnamefont
  {Tang}}, \bibinfo {author} {\bibfnamefont {L.-F.}\ \bibnamefont {Zhang}},
  \bibinfo {author} {\bibfnamefont {G.-Q.}\ \bibnamefont {Zhang}}, \ and\
  \bibinfo {author} {\bibfnamefont {D.-W.}\ \bibnamefont {Zhang}},\ }\bibfield
  {title} {\emph {\bibinfo {title} {Topological Anderson insulators in
  two-dimensional non-Hermitian disordered systems},\ }}\href {\doibase
  10.1103/PhysRevA.101.063612} {\bibfield  {journal} {\bibinfo  {journal}
  {Phys. Rev. A}\ }\textbf {\bibinfo {volume} {101}},\ \bibinfo {pages}
  {063612} (\bibinfo {year} {2020})}\BibitemShut {NoStop}%
\bibitem [{\citenamefont {Jiang}\ \emph {et~al.}(2019)\citenamefont {Jiang},
  \citenamefont {Lang}, \citenamefont {Yang}, \citenamefont {Zhu},\ and\
  \citenamefont {Chen}}]{Jiang2019}%
  \BibitemOpen
  \bibfield  {author} {\bibinfo {author} {\bibfnamefont {H.}~\bibnamefont
  {Jiang}}, \bibinfo {author} {\bibfnamefont {L.-J.}\ \bibnamefont {Lang}},
  \bibinfo {author} {\bibfnamefont {C.}~\bibnamefont {Yang}}, \bibinfo {author}
  {\bibfnamefont {S.-L.}\ \bibnamefont {Zhu}}, \ and\ \bibinfo {author}
  {\bibfnamefont {S.}~\bibnamefont {Chen}},\ }\bibfield  {title} {\emph
  {\bibinfo {title} {Interplay of non-Hermitian skin effects and Anderson
  localization in nonreciprocal quasiperiodic lattices},\ }}\href {\doibase
  10.1103/PhysRevB.100.054301} {\bibfield  {journal} {\bibinfo  {journal}
  {Phys. Rev. B}\ }\textbf {\bibinfo {volume} {100}},\ \bibinfo {pages}
  {054301} (\bibinfo {year} {2019})}\BibitemShut {NoStop}%
\bibitem [{\citenamefont {Lee}\ and\ \citenamefont {Chan}(2014)}]{Lee2014}%
  \BibitemOpen
  \bibfield  {author} {\bibinfo {author} {\bibfnamefont {T.~E.}\ \bibnamefont
  {Lee}}\ and\ \bibinfo {author} {\bibfnamefont {C.-K.}\ \bibnamefont {Chan}},\
  }\bibfield  {title} {\emph {\bibinfo {title} {Heralded Magnetism in
  Non-Hermitian Atomic Systems},\ }}\href {\doibase 10.1103/PhysRevX.4.041001}
  {\bibfield  {journal} {\bibinfo  {journal} {Phys. Rev. X}\ }\textbf {\bibinfo
  {volume} {4}},\ \bibinfo {pages} {041001} (\bibinfo {year}
  {2014})}\BibitemShut {NoStop}%
\bibitem [{\citenamefont {Yamamoto}\ \emph {et~al.}(2019)\citenamefont
  {Yamamoto}, \citenamefont {Nakagawa}, \citenamefont {Adachi}, \citenamefont
  {Takasan}, \citenamefont {Ueda},\ and\ \citenamefont
  {Kawakami}}]{Yamamoto2019}%
  \BibitemOpen
  \bibfield  {author} {\bibinfo {author} {\bibfnamefont {K.}~\bibnamefont
  {Yamamoto}}, \bibinfo {author} {\bibfnamefont {M.}~\bibnamefont {Nakagawa}},
  \bibinfo {author} {\bibfnamefont {K.}~\bibnamefont {Adachi}}, \bibinfo
  {author} {\bibfnamefont {K.}~\bibnamefont {Takasan}}, \bibinfo {author}
  {\bibfnamefont {M.}~\bibnamefont {Ueda}}, \ and\ \bibinfo {author}
  {\bibfnamefont {N.}~\bibnamefont {Kawakami}},\ }\bibfield  {title} {\emph
  {\bibinfo {title} {Theory of Non-Hermitian Fermionic Superfluidity with a
  Complex-Valued Interaction},\ }}\href {\doibase
  10.1103/PhysRevLett.123.123601} {\bibfield  {journal} {\bibinfo  {journal}
  {Phys. Rev. Lett.}\ }\textbf {\bibinfo {volume} {123}},\ \bibinfo {pages}
  {123601} (\bibinfo {year} {2019})}\BibitemShut {NoStop}%
\bibitem [{\citenamefont {Sun}\ and\ \citenamefont {Kou}(2020)}]{Kou2020}%
  \BibitemOpen
  \bibfield  {author} {\bibinfo {author} {\bibfnamefont {G.}~\bibnamefont
  {Sun}}\ and\ \bibinfo {author} {\bibfnamefont {S.-P.}\ \bibnamefont {Kou}},\
  }\bibfield  {title} {\emph {\bibinfo {title} {Biorthogonal quantum
  criticality in non-Hermitian many-body systems},\ }}\href
  {https://arxiv.org/abs/2009.11183} {\bibfield  {journal} {\bibinfo  {journal}
  {arXiv:2009.11183 [cond-mat.str-el]}\ } (\bibinfo {year} {2020})}\BibitemShut
  {NoStop}%
\bibitem [{\citenamefont {Qi}\ and\ \citenamefont
  {Zhang}(2011)}]{ShouCheng2011}%
  \BibitemOpen
  \bibfield  {author} {\bibinfo {author} {\bibfnamefont {X.-L.}\ \bibnamefont
  {Qi}}\ and\ \bibinfo {author} {\bibfnamefont {S.-C.}\ \bibnamefont {Zhang}},\
  }\bibfield  {title} {\emph {\bibinfo {title} {Topological insulators and
  superconductors},\ }}\href {\doibase 10.1103/RevModPhys.83.1057} {\bibfield
  {journal} {\bibinfo  {journal} {Rev. Mod. Phys.}\ }\textbf {\bibinfo {volume}
  {83}},\ \bibinfo {pages} {1057} (\bibinfo {year} {2011})}\BibitemShut
  {NoStop}%
\bibitem [{\citenamefont {Hasan}\ and\ \citenamefont {Kane}(2010)}]{Kane2010}%
  \BibitemOpen
  \bibfield  {author} {\bibinfo {author} {\bibfnamefont {M.~Z.}\ \bibnamefont
  {Hasan}}\ and\ \bibinfo {author} {\bibfnamefont {C.~L.}\ \bibnamefont
  {Kane}},\ }\bibfield  {title} {\emph {\bibinfo {title} {Colloquium:
  Topological insulators},\ }}\href {\doibase 10.1103/RevModPhys.82.3045}
  {\bibfield  {journal} {\bibinfo  {journal} {Rev. Mod. Phys.}\ }\textbf
  {\bibinfo {volume} {82}},\ \bibinfo {pages} {3045} (\bibinfo {year}
  {2010})}\BibitemShut {NoStop}%
\bibitem [{\citenamefont {Zhang}\ \emph {et~al.}(2019)\citenamefont {Zhang},
  \citenamefont {Zhu}, \citenamefont {Zhao}, \citenamefont {Yan},\ and\
  \citenamefont {Zhu}}]{Zhang2019}%
  \BibitemOpen
  \bibfield  {author} {\bibinfo {author} {\bibfnamefont {D.-W.}\ \bibnamefont
  {Zhang}}, \bibinfo {author} {\bibfnamefont {Y.-Q.}\ \bibnamefont {Zhu}},
  \bibinfo {author} {\bibfnamefont {Y.~X.}\ \bibnamefont {Zhao}}, \bibinfo
  {author} {\bibfnamefont {H.}~\bibnamefont {Yan}}, \ and\ \bibinfo {author}
  {\bibfnamefont {S.-L.}\ \bibnamefont {Zhu}},\ }\bibfield  {title} {\emph
  {\bibinfo {title} {Topological quantum matter with cold atoms},\ }}\href
  {\doibase 10.1080/00018732.2019.1594094} {\bibfield  {journal} {\bibinfo
  {journal} {Advances in Physics}\ }\textbf {\bibinfo {volume} {67}},\ \bibinfo
  {pages} {253} (\bibinfo {year} {2019})}\BibitemShut {NoStop}%
\bibitem [{\citenamefont {Tan}\ \emph {et~al.}(2019)\citenamefont {Tan},
  \citenamefont {Zhang}, \citenamefont {Yang}, \citenamefont {Chu},
  \citenamefont {Zhu}, \citenamefont {Li}, \citenamefont {Yang}, \citenamefont
  {Song}, \citenamefont {Han}, \citenamefont {Li}, \citenamefont {Dong},
  \citenamefont {Yu}, \citenamefont {Yan}, \citenamefont {Zhu},\ and\
  \citenamefont {Yu}}]{Tan2019}%
  \BibitemOpen
  \bibfield  {author} {\bibinfo {author} {\bibfnamefont {X.}~\bibnamefont
  {Tan}}, \bibinfo {author} {\bibfnamefont {D.-W.}\ \bibnamefont {Zhang}},
  \bibinfo {author} {\bibfnamefont {Z.}~\bibnamefont {Yang}}, \bibinfo {author}
  {\bibfnamefont {J.}~\bibnamefont {Chu}}, \bibinfo {author} {\bibfnamefont
  {Y.-Q.}\ \bibnamefont {Zhu}}, \bibinfo {author} {\bibfnamefont
  {D.}~\bibnamefont {Li}}, \bibinfo {author} {\bibfnamefont {X.}~\bibnamefont
  {Yang}}, \bibinfo {author} {\bibfnamefont {S.}~\bibnamefont {Song}}, \bibinfo
  {author} {\bibfnamefont {Z.}~\bibnamefont {Han}}, \bibinfo {author}
  {\bibfnamefont {Z.}~\bibnamefont {Li}}, \bibinfo {author} {\bibfnamefont
  {Y.}~\bibnamefont {Dong}}, \bibinfo {author} {\bibfnamefont {H.-F.}\
  \bibnamefont {Yu}}, \bibinfo {author} {\bibfnamefont {H.}~\bibnamefont
  {Yan}}, \bibinfo {author} {\bibfnamefont {S.-L.}\ \bibnamefont {Zhu}}, \ and\
  \bibinfo {author} {\bibfnamefont {Y.}~\bibnamefont {Yu}},\ }\bibfield
  {title} {\emph {\bibinfo {title} {Experimental Measurement of the Quantum
  Metric Tensor and Related Topological Phase Transition with a Superconducting
  Qubit},\ }}\href {\doibase 10.1103/PhysRevLett.122.210401} {\bibfield
  {journal} {\bibinfo  {journal} {Phys. Rev. Lett.}\ }\textbf {\bibinfo
  {volume} {122}},\ \bibinfo {pages} {210401} (\bibinfo {year}
  {2019})}\BibitemShut {NoStop}%
\bibitem [{\citenamefont {Wang}\ \emph {et~al.}(2019)\citenamefont {Wang},
  \citenamefont {Lu}, \citenamefont {Mei}, \citenamefont {Gao}, \citenamefont
  {Li}, \citenamefont {Tang}, \citenamefont {Zhu}, \citenamefont {Jia},\ and\
  \citenamefont {Jin}}]{Wang2019}%
  \BibitemOpen
  \bibfield  {author} {\bibinfo {author} {\bibfnamefont {Y.}~\bibnamefont
  {Wang}}, \bibinfo {author} {\bibfnamefont {Y.-H.}\ \bibnamefont {Lu}},
  \bibinfo {author} {\bibfnamefont {F.}~\bibnamefont {Mei}}, \bibinfo {author}
  {\bibfnamefont {J.}~\bibnamefont {Gao}}, \bibinfo {author} {\bibfnamefont
  {Z.-M.}\ \bibnamefont {Li}}, \bibinfo {author} {\bibfnamefont
  {H.}~\bibnamefont {Tang}}, \bibinfo {author} {\bibfnamefont {S.-L.}\
  \bibnamefont {Zhu}}, \bibinfo {author} {\bibfnamefont {S.}~\bibnamefont
  {Jia}}, \ and\ \bibinfo {author} {\bibfnamefont {X.-M.}\ \bibnamefont
  {Jin}},\ }\bibfield  {title} {\emph {\bibinfo {title} {Direct Observation of
  Topology from Single-Photon Dynamics},\ }}\href {\doibase
  10.1103/PhysRevLett.122.193903} {\bibfield  {journal} {\bibinfo  {journal}
  {Phys. Rev. Lett.}\ }\textbf {\bibinfo {volume} {122}},\ \bibinfo {pages}
  {193903} (\bibinfo {year} {2019})}\BibitemShut {NoStop}%
\bibitem [{\citenamefont {Zhu}\ \emph {et~al.}(2013)\citenamefont {Zhu},
  \citenamefont {Wang}, \citenamefont {Chan},\ and\ \citenamefont
  {Duan}}]{Zhu2013}%
  \BibitemOpen
  \bibfield  {author} {\bibinfo {author} {\bibfnamefont {S.-L.}\ \bibnamefont
  {Zhu}}, \bibinfo {author} {\bibfnamefont {Z.-D.}\ \bibnamefont {Wang}},
  \bibinfo {author} {\bibfnamefont {Y.-H.}\ \bibnamefont {Chan}}, \ and\
  \bibinfo {author} {\bibfnamefont {L.-M.}\ \bibnamefont {Duan}},\ }\bibfield
  {title} {\emph {\bibinfo {title} {Topological Bose-Mott Insulators in a
  One-Dimensional Optical Superlattice},\ }}\href {\doibase
  10.1103/PhysRevLett.110.075303} {\bibfield  {journal} {\bibinfo  {journal}
  {Phys. Rev. Lett.}\ }\textbf {\bibinfo {volume} {110}},\ \bibinfo {pages}
  {075303} (\bibinfo {year} {2013})}\BibitemShut {NoStop}%
\bibitem [{\citenamefont {Zhang}\ and\ \citenamefont
  {Song}(2013{\natexlab{a}})}]{ZhangXZ2013}%
  \BibitemOpen
  \bibfield  {author} {\bibinfo {author} {\bibfnamefont {X.~Z.}\ \bibnamefont
  {Zhang}}\ and\ \bibinfo {author} {\bibfnamefont {Z.}~\bibnamefont {Song}},\
  }\bibfield  {title} {\emph {\bibinfo {title} {Non-Hermitian anisotropic $XY$
  model with intrinsic rotation-time-reversal symmetry},\ }}\href {\doibase
  10.1103/PhysRevA.87.012114} {\bibfield  {journal} {\bibinfo  {journal} {Phys.
  Rev. A}\ }\textbf {\bibinfo {volume} {87}},\ \bibinfo {pages} {012114}
  (\bibinfo {year} {2013}{\natexlab{a}})}\BibitemShut {NoStop}%
\bibitem [{\citenamefont {Wang}\ \emph {et~al.}(2020)\citenamefont {Wang},
  \citenamefont {Yang}, \citenamefont {Guo}, \citenamefont {Zhao},\ and\
  \citenamefont {Kou}}]{WangCan2020}%
  \BibitemOpen
  \bibfield  {author} {\bibinfo {author} {\bibfnamefont {C.}~\bibnamefont
  {Wang}}, \bibinfo {author} {\bibfnamefont {M.-L.}\ \bibnamefont {Yang}},
  \bibinfo {author} {\bibfnamefont {C.-X.}\ \bibnamefont {Guo}}, \bibinfo
  {author} {\bibfnamefont {X.-M.}\ \bibnamefont {Zhao}}, \ and\ \bibinfo
  {author} {\bibfnamefont {S.-P.}\ \bibnamefont {Kou}},\ }\bibfield  {title}
  {\emph {\bibinfo {title} {Effective non-Hermitian physics for degenerate
  ground states of a non-Hermitian Ising model with RT symmetry},\ }}\href
  {\doibase 10.1209/0295-5075/128/41001} {\bibfield  {journal} {\bibinfo
  {journal} {{EPL} (Europhysics Letters)}\ }\textbf {\bibinfo {volume} {128}},\
  \bibinfo {pages} {41001} (\bibinfo {year} {2020})}\BibitemShut {NoStop}%
\bibitem [{\citenamefont {Li}\ \emph {et~al.}(2014)\citenamefont {Li},
  \citenamefont {Zhang}, \citenamefont {Zhang},\ and\ \citenamefont
  {Song}}]{LiC2014}%
  \BibitemOpen
  \bibfield  {author} {\bibinfo {author} {\bibfnamefont {C.}~\bibnamefont
  {Li}}, \bibinfo {author} {\bibfnamefont {G.}~\bibnamefont {Zhang}}, \bibinfo
  {author} {\bibfnamefont {X.~Z.}\ \bibnamefont {Zhang}}, \ and\ \bibinfo
  {author} {\bibfnamefont {Z.}~\bibnamefont {Song}},\ }\bibfield  {title}
  {\emph {\bibinfo {title} {Conventional quantum phase transition driven by a
  complex parameter in a non-Hermitian
  $\mathcal{PT}\ensuremath{-}\mathrm{symmetric}$ Ising model},\ }}\href
  {\doibase 10.1103/PhysRevA.90.012103} {\bibfield  {journal} {\bibinfo
  {journal} {Phys. Rev. A}\ }\textbf {\bibinfo {volume} {90}},\ \bibinfo
  {pages} {012103} (\bibinfo {year} {2014})}\BibitemShut {NoStop}%
\bibitem [{\citenamefont {Zhang}\ and\ \citenamefont
  {Song}(2020)}]{ZhangKL2020}%
  \BibitemOpen
  \bibfield  {author} {\bibinfo {author} {\bibfnamefont {K.~L.}\ \bibnamefont
  {Zhang}}\ and\ \bibinfo {author} {\bibfnamefont {Z.}~\bibnamefont {Song}},\
  }\bibfield  {title} {\emph {\bibinfo {title} {Ising chain with topological
  degeneracy induced by dissipation},\ }}\href {\doibase
  10.1103/PhysRevB.101.245152} {\bibfield  {journal} {\bibinfo  {journal}
  {Phys. Rev. B}\ }\textbf {\bibinfo {volume} {101}},\ \bibinfo {pages}
  {245152} (\bibinfo {year} {2020})}\BibitemShut {NoStop}%
\bibitem [{\citenamefont {Nishiyama}(2020{\natexlab{a}})}]{Yoshihiro2020}%
  \BibitemOpen
  \bibfield  {author} {\bibinfo {author} {\bibfnamefont {Y.}~\bibnamefont
  {Nishiyama}},\ }\bibfield  {title} {\emph {\bibinfo {title}
  {Fidelity-susceptibility analysis of the honeycomb-lattice Ising
  antiferromagnet under the imaginary magnetic field},\ }}\href {\doibase
  10.1140/epjb/e2020-10264-5} {\bibfield  {journal} {\bibinfo  {journal} {The
  European Physical Journal B}\ }\textbf {\bibinfo {volume} {93}},\ \bibinfo
  {pages} {174} (\bibinfo {year} {2020}{\natexlab{a}})}\BibitemShut {NoStop}%
\bibitem [{\citenamefont {Nishiyama}(2020{\natexlab{b}})}]{Nishiyama2020}%
  \BibitemOpen
  \bibfield  {author} {\bibinfo {author} {\bibfnamefont {Y.}~\bibnamefont
  {Nishiyama}},\ }\bibfield  {title} {\emph {\bibinfo {title}
  {Imaginary-field-driven phase transition for the 2D Ising antiferromagnet: A
  fidelity-susceptibility approach},\ }}\href {\doibase
  10.1016/j.physa.2020.124731} {\bibfield  {journal} {\bibinfo  {journal}
  {Physica A: Statistical Mechanics and its Applications}\ }\textbf {\bibinfo
  {volume} {555}},\ \bibinfo {pages} {124731} (\bibinfo {year}
  {2020}{\natexlab{b}})}\BibitemShut {NoStop}%
\bibitem [{\citenamefont {Zhang}\ and\ \citenamefont
  {Song}(2013{\natexlab{b}})}]{Zhang2013}%
  \BibitemOpen
  \bibfield  {author} {\bibinfo {author} {\bibfnamefont {X.~Z.}\ \bibnamefont
  {Zhang}}\ and\ \bibinfo {author} {\bibfnamefont {Z.}~\bibnamefont {Song}},\
  }\bibfield  {title} {\emph {\bibinfo {title} {Geometric phase and phase
  diagram for a non-Hermitian quantum $XY$ model},\ }}\href {\doibase
  10.1103/PhysRevA.88.042108} {\bibfield  {journal} {\bibinfo  {journal} {Phys.
  Rev. A}\ }\textbf {\bibinfo {volume} {88}},\ \bibinfo {pages} {042108}
  (\bibinfo {year} {2013}{\natexlab{b}})}\BibitemShut {NoStop}%
\bibitem [{\citenamefont {Tzeng}\ \emph {et~al.}(2021)\citenamefont {Tzeng},
  \citenamefont {Ju}, \citenamefont {Chen},\ and\ \citenamefont
  {Huang}}]{Tzeng2021}%
  \BibitemOpen
  \bibfield  {author} {\bibinfo {author} {\bibfnamefont {Y.-C.}\ \bibnamefont
  {Tzeng}}, \bibinfo {author} {\bibfnamefont {C.-Y.}\ \bibnamefont {Ju}},
  \bibinfo {author} {\bibfnamefont {G.-Y.}\ \bibnamefont {Chen}}, \ and\
  \bibinfo {author} {\bibfnamefont {W.-M.}\ \bibnamefont {Huang}},\ }\bibfield
  {title} {\emph {\bibinfo {title} {Hunting for the non-Hermitian exceptional
  points with fidelity susceptibility},\ }}\href {\doibase
  10.1103/PhysRevResearch.3.013015} {\bibfield  {journal} {\bibinfo  {journal}
  {Phys. Rev. Research}\ }\textbf {\bibinfo {volume} {3}},\ \bibinfo {pages}
  {013015} (\bibinfo {year} {2021})}\BibitemShut {NoStop}%
\bibitem [{\citenamefont {Wimmer}(2012)}]{Wimmer2012}%
  \BibitemOpen
  \bibfield  {author} {\bibinfo {author} {\bibfnamefont {M.}~\bibnamefont
  {Wimmer}},\ }\bibfield  {title} {\emph {\bibinfo {title} {Algorithm 923:
  Efficient Numerical Computation of the Pfaffian for Dense and Banded
  Skew-Symmetric Matrices},\ }}\href {\doibase 10.1145/2331130.2331138}
  {\bibfield  {journal} {\bibinfo  {journal} {ACM Trans. Math. Softw.}\
  }\textbf {\bibinfo {volume} {38}} (\bibinfo {year} {2012}),\
  10.1145/2331130.2331138}\BibitemShut {NoStop}%
\bibitem [{\citenamefont {Brody}(2013)}]{Brody2013}%
  \BibitemOpen
  \bibfield  {author} {\bibinfo {author} {\bibfnamefont {D.~C.}\ \bibnamefont
  {Brody}},\ }\bibfield  {title} {\emph {\bibinfo {title} {Biorthogonal quantum
  mechanics},\ }}\href {\doibase 10.1088/1751-8113/47/3/035305} {\bibfield
  {journal} {\bibinfo  {journal} {Journal of Physics A: Mathematical and
  Theoretical}\ }\textbf {\bibinfo {volume} {47}},\ \bibinfo {pages} {035305}
  (\bibinfo {year} {2013})}\BibitemShut {NoStop}%
\bibitem [{\citenamefont {Sternheim}\ and\ \citenamefont
  {Walker}(1972)}]{Walker1972}%
  \BibitemOpen
  \bibfield  {author} {\bibinfo {author} {\bibfnamefont {M.~M.}\ \bibnamefont
  {Sternheim}}\ and\ \bibinfo {author} {\bibfnamefont {J.~F.}\ \bibnamefont
  {Walker}},\ }\bibfield  {title} {\emph {\bibinfo {title} {Non-Hermitian
  Hamiltonians, Decaying States, and Perturbation Theory},\ }}\href {\doibase
  10.1103/PhysRevC.6.114} {\bibfield  {journal} {\bibinfo  {journal} {Phys.
  Rev. C}\ }\textbf {\bibinfo {volume} {6}},\ \bibinfo {pages} {114} (\bibinfo
  {year} {1972})}\BibitemShut {NoStop}%
\bibitem [{\citenamefont {Matsumoto}\ \emph {et~al.}(2020)\citenamefont
  {Matsumoto}, \citenamefont {Kawabata}, \citenamefont {Ashida}, \citenamefont
  {Furukawa},\ and\ \citenamefont {Ueda}}]{Matsumoto2020}%
  \BibitemOpen
  \bibfield  {author} {\bibinfo {author} {\bibfnamefont {N.}~\bibnamefont
  {Matsumoto}}, \bibinfo {author} {\bibfnamefont {K.}~\bibnamefont {Kawabata}},
  \bibinfo {author} {\bibfnamefont {Y.}~\bibnamefont {Ashida}}, \bibinfo
  {author} {\bibfnamefont {S.}~\bibnamefont {Furukawa}}, \ and\ \bibinfo
  {author} {\bibfnamefont {M.}~\bibnamefont {Ueda}},\ }\bibfield  {title}
  {\emph {\bibinfo {title} {Continuous Phase Transition without Gap Closing in
  Non-Hermitian Quantum Many-Body Systems},\ }}\href {\doibase
  10.1103/PhysRevLett.125.260601} {\bibfield  {journal} {\bibinfo  {journal}
  {Phys. Rev. Lett.}\ }\textbf {\bibinfo {volume} {125}},\ \bibinfo {pages}
  {260601} (\bibinfo {year} {2020})}\BibitemShut {NoStop}%
\bibitem [{\citenamefont {Duan}\ \emph {et~al.}(2003)\citenamefont {Duan},
  \citenamefont {Demler},\ and\ \citenamefont {Lukin}}]{Lukin2003}%
  \BibitemOpen
  \bibfield  {author} {\bibinfo {author} {\bibfnamefont {L.-M.}\ \bibnamefont
  {Duan}}, \bibinfo {author} {\bibfnamefont {E.}~\bibnamefont {Demler}}, \ and\
  \bibinfo {author} {\bibfnamefont {M.~D.}\ \bibnamefont {Lukin}},\ }\bibfield
  {title} {\emph {\bibinfo {title} {Controlling Spin Exchange Interactions of
  Ultracold Atoms in Optical Lattices},\ }}\href {\doibase
  10.1103/PhysRevLett.91.090402} {\bibfield  {journal} {\bibinfo  {journal}
  {Phys. Rev. Lett.}\ }\textbf {\bibinfo {volume} {91}},\ \bibinfo {pages}
  {090402} (\bibinfo {year} {2003})}\BibitemShut {NoStop}%
\bibitem [{\citenamefont {Lee}\ \emph {et~al.}(2013)\citenamefont {Lee},
  \citenamefont {Gopalakrishnan},\ and\ \citenamefont {Lukin}}]{Lukin2013}%
  \BibitemOpen
  \bibfield  {author} {\bibinfo {author} {\bibfnamefont {T.~E.}\ \bibnamefont
  {Lee}}, \bibinfo {author} {\bibfnamefont {S.}~\bibnamefont {Gopalakrishnan}},
  \ and\ \bibinfo {author} {\bibfnamefont {M.~D.}\ \bibnamefont {Lukin}},\
  }\bibfield  {title} {\emph {\bibinfo {title} {Unconventional Magnetism via
  Optical Pumping of Interacting Spin Systems},\ }}\href {\doibase
  10.1103/PhysRevLett.110.257204} {\bibfield  {journal} {\bibinfo  {journal}
  {Phys. Rev. Lett.}\ }\textbf {\bibinfo {volume} {110}},\ \bibinfo {pages}
  {257204} (\bibinfo {year} {2013})}\BibitemShut {NoStop}%
\bibitem [{\citenamefont {Zhu}\ \emph {et~al.}(2007)\citenamefont {Zhu},
  \citenamefont {Wang},\ and\ \citenamefont {Duan}}]{Zhu2007}%
  \BibitemOpen
  \bibfield  {author} {\bibinfo {author} {\bibfnamefont {S.-L.}\ \bibnamefont
  {Zhu}}, \bibinfo {author} {\bibfnamefont {B.}~\bibnamefont {Wang}}, \ and\
  \bibinfo {author} {\bibfnamefont {L.-M.}\ \bibnamefont {Duan}},\ }\bibfield
  {title} {\emph {\bibinfo {title} {Simulation and Detection of Dirac Fermions
  with Cold Atoms in an Optical Lattice},\ }}\href {\doibase
  10.1103/PhysRevLett.98.260402} {\bibfield  {journal} {\bibinfo  {journal}
  {Phys. Rev. Lett.}\ }\textbf {\bibinfo {volume} {98}},\ \bibinfo {pages}
  {260402} (\bibinfo {year} {2007})}\BibitemShut {NoStop}%
\bibitem [{\citenamefont {Joshi}\ \emph {et~al.}(2013)\citenamefont {Joshi},
  \citenamefont {Nissen},\ and\ \citenamefont {Keeling}}]{Jonathan2013}%
  \BibitemOpen
  \bibfield  {author} {\bibinfo {author} {\bibfnamefont {C.}~\bibnamefont
  {Joshi}}, \bibinfo {author} {\bibfnamefont {F.}~\bibnamefont {Nissen}}, \
  and\ \bibinfo {author} {\bibfnamefont {J.}~\bibnamefont {Keeling}},\
  }\bibfield  {title} {\emph {\bibinfo {title} {Quantum correlations in the
  one-dimensional driven dissipative $XY$ model},\ }}\href {\doibase
  10.1103/PhysRevA.88.063835} {\bibfield  {journal} {\bibinfo  {journal} {Phys.
  Rev. A}\ }\textbf {\bibinfo {volume} {88}},\ \bibinfo {pages} {063835}
  (\bibinfo {year} {2013})}\BibitemShut {NoStop}%
\bibitem [{\citenamefont {Chang}\ \emph {et~al.}(2014)\citenamefont {Chang},
  \citenamefont {Jiang}, \citenamefont {Hua}, \citenamefont {Yang},
  \citenamefont {Wen}, \citenamefont {Jiang}, \citenamefont {Li}, \citenamefont
  {Wang},\ and\ \citenamefont {Xiao}}]{Xiaomin2014}%
  \BibitemOpen
  \bibfield  {author} {\bibinfo {author} {\bibfnamefont {L.}~\bibnamefont
  {Chang}}, \bibinfo {author} {\bibfnamefont {X.}~\bibnamefont {Jiang}},
  \bibinfo {author} {\bibfnamefont {S.}~\bibnamefont {Hua}}, \bibinfo {author}
  {\bibfnamefont {C.}~\bibnamefont {Yang}}, \bibinfo {author} {\bibfnamefont
  {J.}~\bibnamefont {Wen}}, \bibinfo {author} {\bibfnamefont {L.}~\bibnamefont
  {Jiang}}, \bibinfo {author} {\bibfnamefont {G.}~\bibnamefont {Li}}, \bibinfo
  {author} {\bibfnamefont {G.}~\bibnamefont {Wang}}, \ and\ \bibinfo {author}
  {\bibfnamefont {M.}~\bibnamefont {Xiao}},\ }\bibfield  {title} {\emph
  {\bibinfo {title} {Parity-time symmetry and variable optical isolation in
  active-passive-coupled microresonators},\ }}\href {\doibase
  10.1038/nphoton.2014.133} {\bibfield  {journal} {\bibinfo  {journal} {Nature
  Photonics}\ }\textbf {\bibinfo {volume} {8}},\ \bibinfo {pages} {524}
  (\bibinfo {year} {2014})}\BibitemShut {NoStop}%
\bibitem [{\citenamefont {Xing}\ \emph {et~al.}(2017)\citenamefont {Xing},
  \citenamefont {Qi}, \citenamefont {Cao}, \citenamefont {Wang}, \citenamefont
  {Bai}, \citenamefont {Wang}, \citenamefont {Zhu},\ and\ \citenamefont
  {Zhang}}]{Shou2017}%
  \BibitemOpen
  \bibfield  {author} {\bibinfo {author} {\bibfnamefont {Y.}~\bibnamefont
  {Xing}}, \bibinfo {author} {\bibfnamefont {L.}~\bibnamefont {Qi}}, \bibinfo
  {author} {\bibfnamefont {J.}~\bibnamefont {Cao}}, \bibinfo {author}
  {\bibfnamefont {D.-Y.}\ \bibnamefont {Wang}}, \bibinfo {author}
  {\bibfnamefont {C.-H.}\ \bibnamefont {Bai}}, \bibinfo {author} {\bibfnamefont
  {H.-F.}\ \bibnamefont {Wang}}, \bibinfo {author} {\bibfnamefont {A.-D.}\
  \bibnamefont {Zhu}}, \ and\ \bibinfo {author} {\bibfnamefont
  {S.}~\bibnamefont {Zhang}},\ }\bibfield  {title} {\emph {\bibinfo {title}
  {Spontaneous $\mathcal{PT}$-symmetry breaking in non-Hermitian coupled-cavity
  array},\ }}\href {\doibase 10.1103/PhysRevA.96.043810} {\bibfield  {journal}
  {\bibinfo  {journal} {Phys. Rev. A}\ }\textbf {\bibinfo {volume} {96}},\
  \bibinfo {pages} {043810} (\bibinfo {year} {2017})}\BibitemShut {NoStop}%
\end{thebibliography}%

\end{document}